\documentclass[journal]{IEEEtran}
\usepackage[top=0.7in, bottom=0.98in, left=0.7in, right=0.7in]{geometry}
\IEEEoverridecommandlockouts
\usepackage{cite}
\usepackage{subfig}
\usepackage{amsmath,amssymb,amsfonts}
\usepackage{algorithmic}
\usepackage{graphicx}
\usepackage{textcomp}
\usepackage{xcolor}
\usepackage{float}
\usepackage{bm}
\newtheorem{remark}{\rm{\textbf{Remark}}}
\newtheorem{proposition}{\rm{\textbf{Proposition}}}

\newtheorem{theorem}{\rm{\textbf{Theorem}}}
\usepackage{booktabs}   
\usepackage{multirow}   
\usepackage{caption}
\DeclareSubrefFormat{parens}{#1(#2)}
\def\BibTeX{{\rm B\kern-.05em{\sc i\kern-.025em b}\kern-.08em
    T\kern-.1667em\lower.7ex\hbox{E}\kern-.125emX}}
\usepackage{cmap} 
\usepackage[T1]{fontenc}
\usepackage{times} 
\usepackage[utf8]{inputenc}
\usepackage{threeparttable} 
\makeatother

\begin{document}

\title{Unified Generalization for Frequency-Domain Channel Extrapolation Across Near-Field and Far-Field Scenarios}

\author{Haoyu Wang,~\IEEEmembership{Graduate Student Member,~IEEE,}~Zhi Sun,~\IEEEmembership{Senior Member,~IEEE,} \\
Shuangfeng Han,~\IEEEmembership{Senior Member,~IEEE,}~Xiaoyun Wang,~and Zhaocheng Wang,~\IEEEmembership{Fellow,~IEEE}

    \thanks{Haoyu Wang, Zhi Sun, and Zhaocheng Wang are with the Department of Electronic Engineering, Tsinghua University, Beijing 100084 China (e-mail: wanghy22@mails.tsinghua.edu.cn; zhisun@ieee.org; zcwang@tsinghua.edu.cn).}
    \thanks{Shuangfeng Han and Xiaoyun Wang are with the China Mobile Research Institute, Beijing 100053, China. (e-mail: hanshuangfeng@chinamobile.com; wangxiaoyun@chinamobile.com)}
    \thanks{This work was supported by the National Key R\&D Program of China under Grant 2022YFB2902004.}
    \thanks{Corresponding Author: Zhi Sun.}
}

\maketitle
\begin{abstract}
With the increasing number of antennas, the near-field effect becomes non-negligible in large-scale multiple-input multiple-output (MIMO) systems. Accurate and low-overhead channel acquisition is therefore essential to maintain system performance in both far-field and near-field scenarios. Deep learning (DL)-based frequency-domain channel extrapolation offers an effective solution for achieving low-overhead and accurate channel acquisition. However, due to the challenges posed by distance-dependency and environment-dependency, existing DL extrapolators fail to simultaneously generalize to unseen near-field and far-field channels, hindering their practical deployment. In this paper, we propose a physically interpretable framework that unifies the generalization capability of frequency-domain DL extrapolators across near-field and far-field scenarios. Firstly, we propose the key insights of heterogeneous angular profile and unified delay profile for unified generalization. Explicitly, the angular profiles are heterogeneous across near-field and far-field scenarios, while the delay profiles exhibit a unified sparsity pattern that can be aligned. Secondly, we introduce a physics-based disentanglement and alignment framework consisting of three steps: multi-cluster decoupling, angle-delay feature disentanglement, and delay-domain alignment. This allows the DL model to learn distributionally stable delay-domain features while directly reusing heterogeneous angular-domain features. Thirdly, we design a \underline{U}nified \underline{N}ear-f\underline{i}eld and \underline{F}ar-f\underline{i}eld \underline{DL} \underline{E}xtrapolator (UNiFi-DLE) and elaborate its workflow, including dataset preparation, model training, and inference. Extensive simulations and sim-to-real experiments validate that the proposed UNiFi-DLE robustly generalizes across both unseen near-field and far-field scenarios, consistently outperforming state-of-the-art methods.
\end{abstract}


\section{Introduction}
\subsection{Background}
\begin{figure*}[t]
        \centering
        \includegraphics[width=1\textwidth]{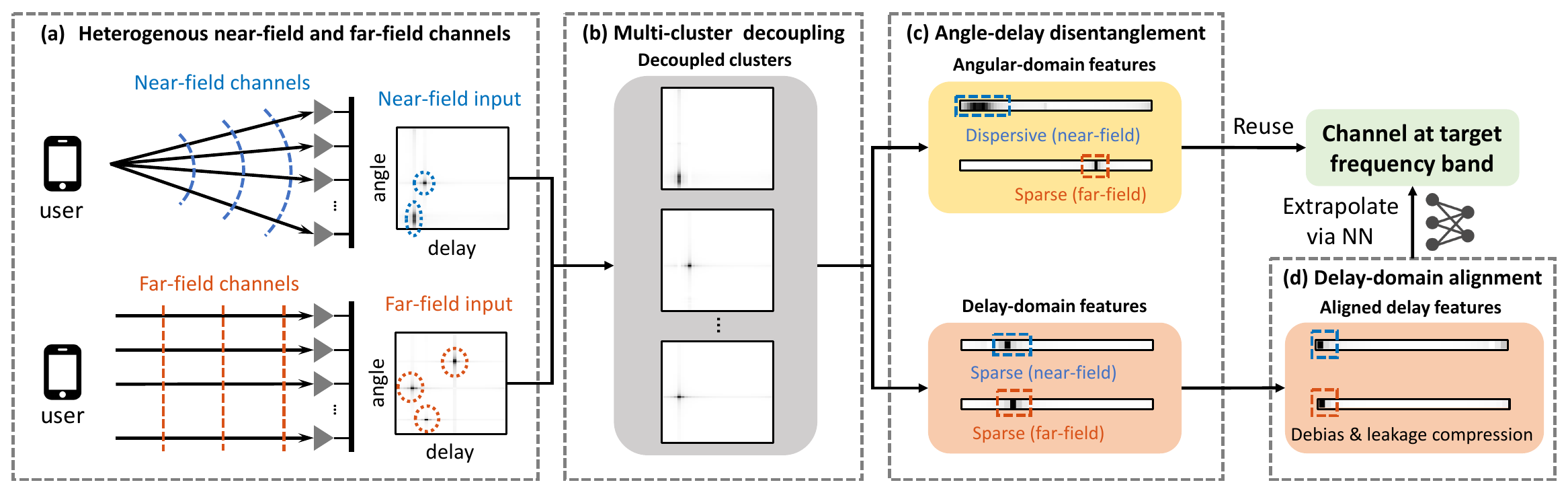}
    \captionsetup{font=footnotesize}
    \caption{(a) Heterogeneous electromagnetic propagation across near-field and far-field scenarios. (b)-(d) Proposed physics-based disentanglement and alignment framework to address the heterogeneity between near-field and far-field channels, which yields the heterogeneous angular-domain features and aligned delay-domain features.}
    \label{fig: framework}
    \vspace{-15pt}
\end{figure*}
Multiple-input multiple-output (MIMO) stands as a pivotal technology for enhancing the spectral efficiency and capacity of wireless communication systems. With a large number of antennas, the large-scale MIMO arrays introduce significant multiplexing and diversity gains. Fully realizing these gains, however, necessitates accurate acquisition of channel state information (CSI) \cite{twc_multicell_2019_Shafin}. Moreover, to ensure these theoretical gains translate into effective spectral efficiency, the overhead associated with CSI acquisition should be minimized \cite{wc_zeng_2021_towad}.

The increasing aperture of antenna arrays introduces a paradigm shift in electromagnetic wave propagation, where near-field and far-field regions coexist \cite{network_new_2025_wang}. This shift poses significant challenges for accurate and low-overhead CSI acquisition in large-scale MIMO systems. Conventionally, the wavefronts of the electromagnetic waves are assumed to be planar  \cite{twc_he_2023_Beamspace,tcom_Deep_2023_ma,tcom_deep_2021_ma}. Then, the MIMO channel exhibits obvious sparsity in the angular-domain and motivates various accurate and low-overhead channel acquisition algorithms, e.g., the compressed sensing-based algorithms \cite{tsp_cs_2015_rao}. However, such an approximation is only valid in the far-field scenarios and is invalid in the near-field regions, where the wavefront of the propagated electromagnetic wave is spherical. Then, the cross-antenna channel correlation is not only determined by the propagation directions, but also affected by the distances \cite{jsac_2025_far_miao,twc_2024_near_cui}. Consequently, the angular-domain sparse nature of the large-scale MIMO channel is not held in the near-field scenarios, which degrades the accuracy and efficiency of the conventional CSI acquisition algorithms designed under far-field assumptions. Based on electromagnetic wave propagation theory, the boundary between near-field and far-field is determined by the Rayleigh distance, which scales quadratically with the array aperture \cite{apm_Fraunhofer_2017_Selvan}. When the aperture reaches the scale of meters, the near-field region can achieve tens or even hundreds of meters \cite{jsac_2025_far_miao,twc_2024_near_cui}. Thus, how to achieve unified and efficient CSI acquisition across near-field and far-field scenarios is vital and of practical significance to provide seamless service in large-scale massive MIMO systems \cite{twc_2022_channel_cui,wcl_unifying_2024_chen,network_new_2025_wang}. 

Frequency-domain channel extrapolation serves as a unified CSI acquisition approach to reduce the pilot overhead in large-scale MIMO systems, which is based on the frequency-invariant geometrical path parameters across near-field and far-field scenarios \cite{jstsp_Han_2019_tracking,twc_2022_channel_cui}. Based on the electromagnetic wave propagation theory, paths in the far-field and near-field share the same parameters, which implies that the channel responses at different subcarriers inherently exhibit the same underlying physical structure. Thus, the channel at a target frequency band can be extrapolated from the measured band. The frequency-domain dependency, governed by complex electromagnetic propagation, is inherently non-linear and high-dimensional. Conventional model-based extrapolation often relies on explicit parameter estimation, whose performance is affected by the noise and inter-path interference. This gap motivates the adoption of deep learning (DL), which excels at approximating non-linear dependency. Therefore, numerous neural network structures have been proposed to enhance the channel extrapolation accuracy \cite{Asilomar_Alrabeiah_2019_deep,cl_time_2025_pang,openj_jiang_2020_deep,wcl_Yao_2024_loss}.

\subsection{Research Challenge}
In the co-existence of near-field and far-field effects, unified frequency-domain channel extrapolation encounters two fundamental challenges as follows. 
\begin{itemize}
    \item \textbf{Distance-dependency:} Owing to the near-field effect, the channels are not only determined by the parameters of angle of departure/arrival, delay, and complex gain, but also affected by the distance parameter. Thus, the degrees of freedom of the large-scale MIMO channel distribution are further increased in the co-existence of near-field and far-field effects. 
    \item \textbf{Environment-dependency:} Based on the geometrical channel model, the propagated paths interact with the user, base station (BS), and scatterers in the environment, which determines the multipath parameters. Owing to the diversity of the wireless environment, the distribution of both near-field and far-field channels is highly diverse across different environments \cite{wang2025path}. 
\end{itemize}
Consequently, out-of-distribution (OOD) channel samples are inevitable across far-field and near-field scenarios, which fundamentally limits the unified generalizability.

\subsection{Motivation and Our Contributions}
In our previous work, the path-oriented DL extrapolator with path alignment (PO-DLE+PA) was proposed to partly address the environment-dependency of large-scale MIMO channel in the far-field scenarios \cite{wang2025generalizable}. Nevertheless, the PO-DLE+PA heavily relies on the angular-delay domain joint sparsity of the extracted responses, which is severely deteriorated owing to the distance-dependency. To the best of our knowledge, there is no existing work to achieve unified generalization for frequency-domain channel extrapolation across diverse near-field and far-field scenarios. In this paper, we aim to achieve unified generalizability in frequency-domain DL extrapolators, which simultaneously resolves the environment-dependency and distance-dependency. Then, the trained DL-extraplator can directly generalize to unseen diverse near-field and far-field scenarios without domain adaptation, where the data collection cost for practical deployment can be greatly reduced. Our specific contributions can be summarized as follows:
\begin{itemize}
    \item Firstly, we propose two key insights for unified generalizability, namely, the heterogeneous angular profile and unified delay profile. As depicted in Fig.~\ref{fig: framework}(a), the clusters in the near-field and far-field exhibit heterogeneous angular profiles, which is the bottleneck for unified generalizability. Under planar wavefront propagation, the angular profile of a far-field cluster is sparse, whereas its near-field counterpart becomes dispersive as a result of the spherical wavefront. Consequently, the heterogeneous angular-profile will directly degrade the distribution alignment for generalizability enhancement. In contrast, we prove that the delay-profile is unified across near-field and far-field scenarios. Explicitly, the delay-profiles of both near-field and far-field clusters are sparse, and independent of the specific propagation wavefronts.
    \item Secondly, as shown in Fig.~\ref{fig: framework}(b)-(d), a novel physics-based disentanglement and alignment framework is proposed based on the key insights, which is composed of the multi-cluster decoupling, angle-delay disentanglement, and delay-domain alignment. We verify that both the near-field and far-field clusters share the rank-one and mutual orthogonal properties, and a singular value decomposition (SVD)-based multi-cluster decoupling is adopted. Then, angular-delay disentanglement is proposed to yield the heterogeneous angular-domain features and unified delay-domain features. Supported by the evidence from real-world channel measurements, we prove that the angular-domain features are frequency-independent and can be directly reused at the target frequency band. The delay-domain features are frequency-dependent and can be learned by neural networks. Further, the delay-domain alignment is applied to debias and compress leakage of the delay-domain features across various environments, which guarantees the model generalizability.
    \item Thirdly, a \underline{U}nified \underline{N}ear-f\underline{i}eld and \underline{F}ar-f\underline{i}eld \underline{DL} \underline{E}xtrapolator (UNiFi-DLE) is proposed to achieve the unified generalizability, where the dedicated workflow is designed. In the preparation of the training dataset, we enforce frequency-domain consistency. Explicitly, the label is constructed as the aligned delay-domain features from the target band, ensuring it retains the proper correlation with the aligned input features from the measured band. This theoretically allows the model to learn the valid cross-frequency mapping. Then, the model training and inference pipelines are designed to leverage the physics-based disentanglement and alignment.
    \item Extensive simulations and sim-to-real experiments are conducted to justify the proposed UNiFi-DLE. Compared to the state-of-the-art, the proposed UNiFi-DLE can simultaneously achieve the best generalizability in both unseen far-field and near-field scenarios. Specifically, the generalizability gain of UNiFi-DLE becomes more obvious as the distance decreases. In the sim-to-real experiments, the unified generalizability of the proposed UNiFi-DLE is also robustly achieved, which further validates its feasibility in real-world deployment. 
\end{itemize}

\subsection{Notations}
$\mathbb{C}^{m\times n}$ denotes the space of $m\times n$ complex matrices; $(\cdot)^T$, $(\cdot)^H$, $\text{conj}(\cdot)$ denote the transpose, Hermitian transpose, and conjugate operations; $|\cdot|$, $\Vert\cdot\Vert_2$ and $\Vert\cdot\Vert_F$ stand for the absolute value of a scalar, the $l_2$ norm of a vector and Frobenius norm of a matrix, respectively; $\mathbb{E}\{\cdot\}$ denotes the statistical expectation; $\odot$ denotes the element-wise Hadamard product; $U(a,b)$ denotes the uniform distribution between $a$ and $b$. 
\section{Related Works}
\subsection{Generalizable Learning for Frequency-Domain Channel Extrapolation}
In \cite{twc_liu_2024_deep,chinacom_han_2024_AI,jstsp_guo_2022_user}, data augmentation techniques, including angular-delay domain shift (ADS), flip, and random phase shift (RPS), have been proposed to enhance the diversity of the training dataset. However, random data augmentation cannot fully address the diverse distribution shift under heterogeneous propagation. In \cite{mobicom_Banerjee_2024_HORCRUX}, a hierarchical neural network, HORCRUX, is designed to robustly give initial guesses for parameter optimization. However, channel distribution shift will affect the quality of the delay guesses, and the channel correlation between different antennas is not leveraged. In our previous work \cite{wang2025generalizable}, PO-DLE+PA is proposed to partly address the distribution shift of the large-scale MIMO channel in the far-field scenarios. In the near-field region, the power dispersion will severely affect the distribution alignment performance, where the model generalizability will obviously degrade. 
\subsection{Unified Near-Field and Far-Field Communications}
Unified near-field and far-field communications aim to develop a unified framework that can seamlessly serve the users across near-field and far-field regions, which is free of the user position knowledge. In \cite{wcl_unifying_2024_chen}, a modified Fourier plane-wave (FPW) series approximation is proposed to accommodate the presence of near-field and far-field. In \cite{network_new_2025_wang}, a unified codebook is proposed based on K-SVD, which serves as a unified measurement matrix for near-field and far-field users. In \cite{twc_wavenumber_2025_weng}, a wavenumber-domain unified beam training framework is proposed, which interprets the spherical waves as the superposition of multiple planar waves. 
\section{Problem Formulation and Key Solution}
In this section, we first formulate the unified frequency-domain extrapolation across near-field and far-field in Sec.~\ref{subsec: problem}. To solve the unified generalization problem, two key insights are derived in Sec.~\ref{subsec: challenge} and Sec.~\ref{subsec: key insight}, respectively. Thereafter, the key solution with intuitive physics interpretation is proposed in Sec.~\ref{subsec: physics-ddaa}.

\subsection{Problem Formulation: Unified Generalization for Frequency-Domain Channel Extrapolation}
\label{subsec: problem}
In this paper, we consider a multi-carrier large-scale MIMO system serving a single-antenna user, where the number of antennas at the BS is denoted as $N_{\rm T}$. Assume the non-overlapping measured frequency band $\mathcal{B}^{(\rm m)}=\{f^{(\rm m)}_i\}_{i=1}^{K^{(\rm m)}}$ and the target frequency band $\mathcal{B}^{(\rm e)}=\{f^{(\rm e)}_{i}\}_{i=1}^{K^{(\rm e)}}$, where $f^{(\rm m)}_{i}$ and $f^{(\rm e)}_i$ denote the $i$th subcarrier in the measured and target frequency bands, respectively. Denote the channel response at frequency $f$ as $\mathbf{h}(f)\in\mathbb{C}^{N_{\rm T}\times 1}$. Thus, the frequency-domain channel extrapolation aims to construct the mapping from the channel in the measured frequency band $\mathcal{B}^{(\rm m)}$ to the target frequency band $\mathcal{B}^{(\rm e)}$, i.e., 
\begin{equation}
    \label{equ: channel extrapolation}
    \psi:\mathbb{C}^{N_{\rm T}\times K_{\rm m}}\to\mathbb{C}^{N_{\rm T}\times K_{\rm e}}, ~ {\mathbf{H}^{(\rm m)}}\mapsto\mathbf{H}^{(\rm e)}=\psi(\mathbf{H}^{(\rm m)}),
\end{equation}
where $\mathbf{H}^{(x)}=\big[\mathbf{h}\big(f_{1}^{(x)}\big),\ldots,\mathbf{h}\big(f_{{K}^{(x)}}^{(x)}\big)\big]\in\mathbb{C}^{N_{\rm T}\times K^{(x)}}$, and indicator $x\in\{{\rm m}, {\rm e}\}$. To approximate the target mapping $\psi$, the DL channel extrapolator is usually adopted. Denote the training dataset as $\mathcal{D}=\big\{\big(\mathbf{H}^{(\rm m)}_{i},\mathbf{H}^{(\rm e)}_i\big)|i=1,2,\ldots,M\big\}$, where $\big(\mathbf{H}^{(\rm m)}_{i},\mathbf{H}^{(\rm e)}_i\big)$ denotes the $i$th input-label sample in the dataset and $M$ denotes the dataset size. Then, the DL-channel extrapolator can be optimized by minimizing the mean square error (MSE) loss over $\mathcal{D}$. 

Heterogeneous wave propagation occurs when the users are located in different positions in the large-scale MIMO systems, which is plotted in Fig.~\ref{fig: framework}(a). Based on the electromagnetic wave propagation theory, the far-field and near-field regions of a large-scale MIMO system can be distinguished with the Rayleigh distance $Z=2D^2/\lambda$, where $D$ denotes the array aperture and $\lambda$ denotes the wavelength \cite{apm_Fraunhofer_2017_Selvan}. Explicitly, when the users and the last-hop scatterers are in the far-filed of the large-scale MIMO system, the propagated wavefronts are planar. Then, the channel $\mathbf{h}^{\rm(FF)}(f)$ in the far-field can be formulated as \cite{tcom_2024_deep_ma} 
\begin{equation}
    \label{equ: channel model far}
    \mathbf{h}^{\rm {(FF)}}(f)=\sum_{l=1}^{N_{\rm cl}}\sum_{i=1}^{N_{{\rm r,}i}}\alpha_{l,i}e^{-{\rm j} 2\pi f\tau_{l,i}}\mathbf{a}^{(\rm FF)}(\varphi_{l,i},\theta_{l,i}),
\end{equation}
where $N_{\rm cl}$ denotes the number of physical clusters, $N_{{\rm r},i}$ denotes the number of paths within $i$th cluster, $\alpha_{l,i},\tau_{l,i}, \varphi_{l,i}, \theta_{l,i}$ denote the complex gain, delay, azimuth angle of departure, elevation angle of departure of the $i$th path in the $l$th cluster. Under the planar wave propagation, the $n$th element in the steering vector $\mathbf{a}^{(\rm FF)}(\varphi,\theta)$ can be calculated by 
\begin{equation}
    \label{equ: steering far}
    \big[\mathbf{a}^{(\rm FF)}(\varphi,\theta)\big]_{n}=e^{{\rm j}\kappa \mathbf{r}(\varphi,\theta)^{T}\mathbf{d}_{n}},
\end{equation}
where $\kappa=2\pi f/c$ denotes the wavenumber ($c$ denotes the lightspeed), $\mathbf{r}(\varphi,\theta)=[\sin(\theta)\cos(\varphi),\sin(\theta)\sin(\varphi),\cos(\theta)]^T$ denotes the wavefront direction, and $\mathbf{d}_{n}$ denotes the location vector of the $n$th antenna element in the array. When the users or the last-hop scatterers are in the near-field region, the propagated wavefronts are transformed into spherical. In this case, the channel $\mathbf{h}^{{\rm (NF)}}(f)$ in the near-field can be represented by \cite{twc_2022_channel_cui,jsac_2025_far_miao}
\begin{equation}
    \label{equ: channel model near}
    \mathbf{h}^{\rm {(NF)}}(f)=\sum_{l=1}^{N_{\rm cl}}\sum_{i=1}^{N_{{\rm r,}i}}\alpha_{l,i}e^{-{\rm j} 2\pi f\tau_{l,i}}\mathbf{a}^{(\rm NF)}(\mathbf{x}_{l,i}),
\end{equation}
where $\mathbf{x}_{l,i}$ denotes the location of the scatter or the user, and the $n$th element in the steering vector $\mathbf{a}^{(\rm NF)}(\mathbf{x})$ can be calculated by 
\begin{equation}
    \label{equ: spherical}
    [\mathbf{a}^{(\rm NF)}(\mathbf{x})]_{n}=e^{-{\rm j}\kappa(\Vert\mathbf{x}-\mathbf{d}_{n}\Vert_2-\Vert\mathbf{x}-\mathbf{d}_{0}\Vert_2)}. 
\end{equation}
Thus, the unified generalization aims to learn a unified model, which is robust and generalizable to extrapolate the channel in unseen near-field and far-field scenarios. 

\subsection{Key Insight 1: Heterogeneous Angular Profile}
\label{subsec: challenge}
The heterogeneity of the planar and spherical wave propagation is the bottleneck for the unified generalization across near-field and far-field scenarios. In our previous study \cite{wang2025generalizable}, the generalizability of the PO-DLE+PA relies on the joint-sparsity of extracted path/cluster responses in the angular-delay domain. Explicitly, the power of a single path or cluster is concentrated in the angular-delay domain to facilitate the path-level/cluster-level alignment. Nevertheless, this assumption cannot be guaranteed when confronted with spherical wave propagation in the near-field. In the heterogeneous far-field and near-field channels, the steering vectors $\mathbf{a}^{(\rm FF)}(\varphi, \theta)$ and $\mathbf{a}^{(\rm NF)}(\mathbf{x})$ exhibit distinct phase structures among antenna elements. For the uniform linear array (ULA) or uniform planar array (UPA), the phase shift in $[\mathbf{a}^{(\rm FF)}(\varphi, \theta)]_{n}$ is a linear function of index $n$, while the counterpart in $[\mathbf{a}^{(\rm NF)}(\mathbf{x})]_{n}$ is a non-linear function \cite{jsac_2025_far_miao}. Owing to the non-linear phase variation among antenna elements, the spherical wave propagation results in a dispersive angular profile of a path or a cluster in the near-field. 

\begin{figure*}[t]
    \captionsetup{font=footnotesize,justification=centering}
    \quad\subfloat[Angular profile 
    ]{\includegraphics[width=0.5\textwidth]{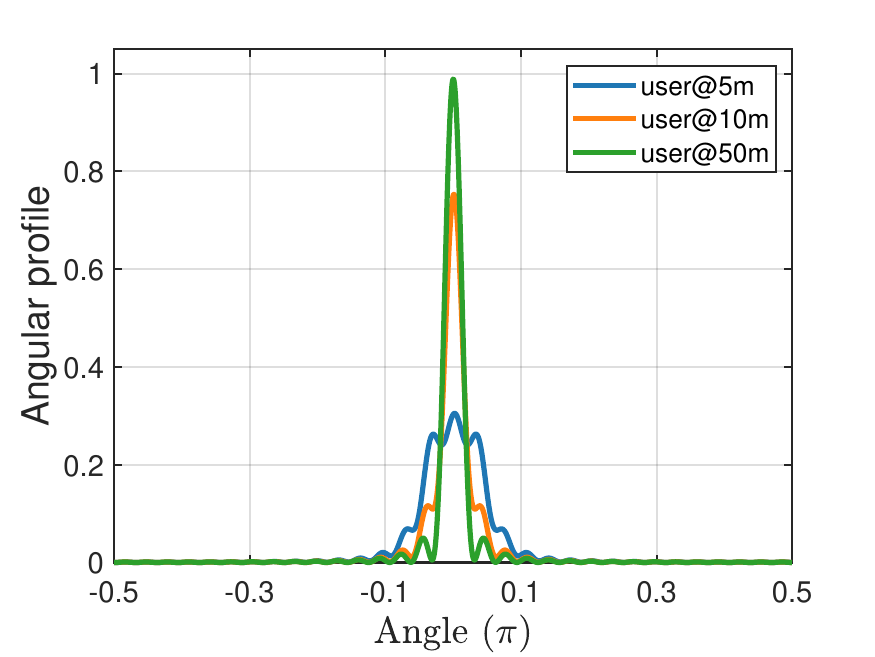}\captionsetup{font=footnotesize,justification=centering}\label{subfig: angular profile}}
    \subfloat[Delay profile 
    ]{\includegraphics[width=0.5\textwidth]{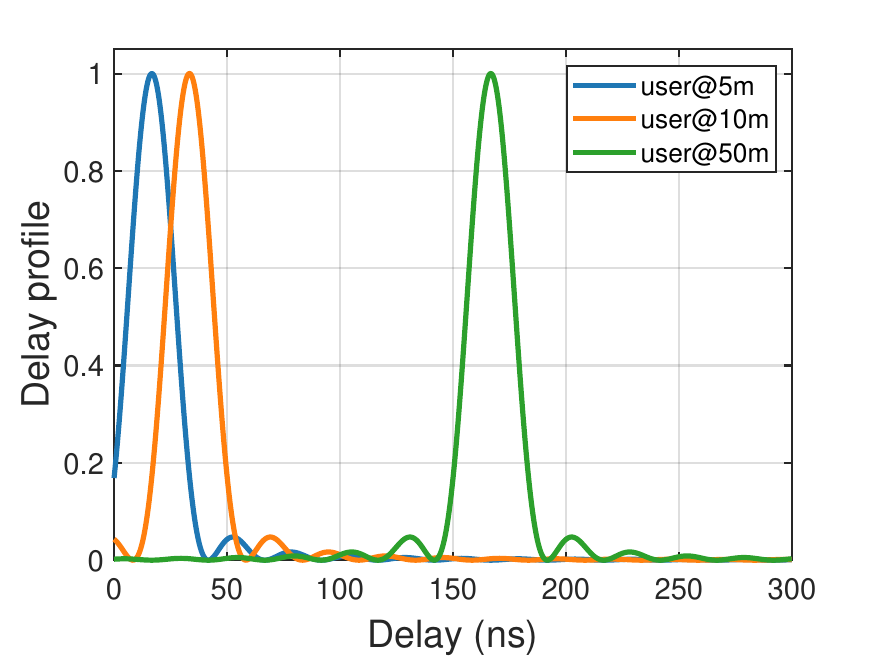}\captionsetup{font=footnotesize,justification=raggedright}\label{subfig: delay profile}}
    \captionsetup{justification=raggedright,singlelinecheck=false,font=small}
    \caption{Normalized power profiles of a LOS path from near-field to far-field, where the Rayleigh distance is 45 m. }
    \label{fig: profile}
    \vspace{-10pt}
\end{figure*}

Consider a numerical example for the understanding of angular-profile heterogeneity. Denote the response of $l$th path as $\mathbf{H}^{(\rm m)}_l\in\mathbb{C}^{N_{\rm T}\times K^{(\rm m)}}$, then the angular profile $P_l^{(\rm a)}(\varphi,\theta)$ can be defined as 
\begin{equation}
    \label{equ: angular profile}
    P_l^{(\rm a)}(\varphi,\theta)=\frac{1}{N_{\rm T}}\left\Vert\big(\mathbf{a}^{(\rm FF)}(\theta,\varphi)\big)^H\mathbf{H}^{(\rm m)}_l\right\Vert_2^2,
\end{equation}
which characterizes the power distribution in different spatial directions. Let the BS be equipped with a 32-antenna half-wavelength ULA operating at 3.4 GHz. The number of subcarriers $K^{(\rm m)}$ is set as 32, and the bandwidth is 40 MHz. Then, the angular profile of the line-of-sight (LOS) path when the user-BS distance varies from 5 m to 50 m is shown in Fig.~\ref{fig: profile}\subref{subfig: angular profile}. In the far-field region (user@50m), the power distribution of the LOS path is highly concentrated, and the path response can be easily aligned. Nevertheless, when the user is located in the near-field region (e.g., user@5m and user@10m in Fig.~\ref{fig: profile}\subref{subfig: angular profile}), the angular-domain sparsity of the LOS path is degraded, which becomes dispersive. Meanwhile, the angular spread is distance-dependent, which further complicates the alignment of different paths. Based on the analysis above, the unified generalization for far-field and near-field channel extrapolation is more challenging compared to existing far-field cross-environment generalization. 

\subsection{Key Insight 2: Unified Delay Profile}
\label{subsec: key insight}

Compared to the complicated angular profile, the delay profile is much more stable within a path or cluster across both near-field and far-field scenarios. This stability stems from the fact that propagation delays are primarily determined by geometric lengths, which remain consistent regardless of wavefront curvature.  Without loss of generality, assume the subcarriers in $\mathcal{B}^{(\rm m)}$ and $\mathcal{B}^{(\rm e)}$ have equal spacing $\Delta f$, i.e., $f^{(x)}_{i}=f_{1}^{(x)}+(i-1)\Delta f$ for $x\in\{{\rm m,e}\}$. In both far-field and near-field scenarios, the steering vectors $\mathbf{a}^{(\rm FF)}(\varphi,\theta)$ and $\mathbf{a}^{(\rm FF)}(\mathbf{x})$ can be approximated as frequency independent when the gap between the measured and target frequency band is far less than the carrier frequency, i.e., $\max_{i,j}|f^{(\rm m)}_{i}-f^{(\rm e)}_j|\ll\min\{f_{1}^{(\rm m)}, f^{(\rm e)}_1\}$ \cite{3gpp.38.901,twc_he_2023_Beamspace,twc_2024_near_cui}. Then, the channel matrixes $\mathbf{H}^{(\rm m)}$ and $\mathbf{H}^{(\rm e)}$ in the far-field and near-field can be represented by 
\begin{equation}
    \label{equ: Hm far}
    \mathbf{H}^{(x, {\rm FF})}=\sum_{l=1}^{N_{\rm cl}}\sum_{i=1}^{N_{{\rm r}, i}}\alpha_{l,i}e^{-{\rm j}2\pi f^{(x)}_{1}\tau_{l,i}}\mathbf{a}^{(\rm FF)}(\varphi_{l,i},\theta_{l,i})\mathbf{b}^{T}_{x}(\tau_{l,i}),
\end{equation}
and 
\begin{equation}
    \label{equ: Hm near}
    \mathbf{H}^{(x, {\rm NF})}=\sum_{l=1}^{N_{\rm cl}}\sum_{i=1}^{N_{{\rm r}, i}}\alpha_{l,i}e^{-{\rm j}2\pi f^{(x)}_{1}\tau_{l,i}}\mathbf{a}^{(\rm NF)}(\mathbf{x}_{l,i})\mathbf{b}^{T}_{x}(\tau_{l,i}),
\end{equation}
where $\mathbf{b}_{x}(\tau)$ denotes the frequency-domain response vector in the measured band and can be represented by 
\begin{equation}
    \label{equ: bm}
    \mathbf{b}_{x}(\tau)=\left[1,e^{{\rm -j}2\pi\Delta f\tau},\ldots,e^{{\rm -j}2\pi(K^{(x)}-1)\Delta f\tau}\right]^T,
\end{equation}
for $x\in\{{\rm m, e}\}$. Similar to \eqref{equ: angular profile}, the delay profile $P^{(\rm d)}_l(\tau)$ of $\mathbf{H}^{(\rm m)}_l$ can be defined as 
\begin{equation}
    \label{equ: delay profile}
    P^{(\rm d)}_l(\tau)=\frac{1}{N_{\rm T}}\left\Vert\mathbf{H}^{(\rm m)}_l\text{conj}(\mathbf{b}_{\rm m}(\tau))\right\Vert_2^2.
\end{equation}
We consider the delay profile of the LOS path in the near-field and far-field as an example, where the user location is denoted as $\mathbf{x}_{\rm ue}$ and the complex path gain is denoted as $\alpha_{\rm LOS}$. Then, the delay profile $P^{(\rm d)}_{\rm LOS}(\rm \tau)$ of the LOS path can be calculated 
\begin{equation}
    \label{equ: LOS delay profile}
    P^{(\rm d)}_{\rm LOS}({\rm \tau})=|\alpha_{\rm LOS}|^2\left|\frac{\sin\left({\pi K^{(\rm m)}\Delta f(\tau-\tau_{\rm LOS})}\right)}{\sin{\left(\pi \Delta f(\tau-\tau_{\rm LOS})\right)}}\right|^2,
\end{equation}
where $\tau_{\rm LOS}=\Vert\mathbf{x}_{\rm ue}-\mathbf{d}_{0}\Vert_2/c$. Thus, it can be found that the delay profile is unified for both near-field and far-field scenarios. Also, we consider a numerical example for more intuitive understanding, where the parameter settings are aligned with Fig.~\ref{fig: profile}\subref{subfig: angular profile}. Then, the delay profile of the LOS path from near-field to far-field is illustrated in Fig.~\ref{fig: profile}\subref{subfig: delay profile}. Despite the peak bias, the delay profile of the LOS path shares a unified shape across near-field and far-field regions. Thus, the utilization of the unified delay profile is critical to achieve unified generalization across near-field and far-field.

\subsection{Solution: Physics-based Disentanglement and Alignment}
\label{subsec: physics-ddaa}
Based on the detailed analysis in Sec.~\ref{subsec: challenge} and \ref{subsec: key insight}, the physics-based distribution disentanglement and alignment framework is proposed to achieve the unified generalization across near-field and far-field scenarios, which is plotted in Fig.~\ref{fig: framework}(b)-(d). Since the channels in both near-field and far-field scenarios are composed of multiple clusters \cite{jsac_2025_far_miao}, multi-cluster decoupling $\mathbf{H}^{(\rm m)}$ is first applied to decompose the original channel matrix into the summation of multiple clusters $\mathbf{H}^{(\rm m)}=\sum_{k}\mathbf{C}^{(\rm m)}_{k}$, which is shown in Fig.~\ref{fig: framework}(b). Thus, the complex shift of multi-cluster structure (including the number and dependency of the clusters) across far-field and near-field scenarios can be effectively addressed. Owing to the heterogeneous spherical and planar wave propagation in the near-field and far-field scenarios, the decoupled clusters cannot be effectively aligned due to the angular-profile dispersion in the near-field channels. To this end, the decoupled clusters $\{\mathbf{C}^{(\rm m)}_{k}\}$ are processed by the angle-delay disentanglement module in Fig.~\ref{fig: framework}(c), which generates angular-domain features $\{\mathbf{f}_{k}^{(\rm a)}\}$ and the delay-domain features $\{\mathbf{f}_{k}^{(\rm d,m)}\}$. Note that the disentanglement operation is invertible. In other words, the cluster $\mathbf{C}_{k}^{(\rm m)}$ can be precisely reconstructed from the features $(\mathbf{f}_{k}^{(\rm a)},\mathbf{f}_{k}^{(\rm d,m)})$. Here, angular-domain feature $\mathbf{f}^{(\rm a)}_{k}\in\mathbb{C}^{N_{\rm T}\times1}$ is frequency-independent. On the contrary, the delay-domain feature $\mathbf{f}^{(\rm d,m)}_{k}$ is frequency-dependent and can reflect the delay-domain cluster power distribution. To account for frequency dependency, the trained neural network takes the measured delay-domain feature $\{\mathbf{f}_k^{(\rm d,m)}\}$ as input and generates the extrapolated feature $\{\widehat{\mathbf{f}}_k^{(\rm d,e)}\}$. Then, the channel at the target frequency band can be reconstructed with the angular-domain features $\{\mathbf{f}^{(\rm a)}_{k}\}$ and the extrapolated delay-domain features $\{\widehat{\mathbf{f}}_{k}^{(\rm d,e)}\}$. Since the angular-domain features $\{\mathbf{f}_{k}^{(\rm a)}\}$ are directly reused at the target frequency-band through the disentanglement, the DL extrapolator will not fit the distribution of $\{\mathbf{f}_{k}^{(\rm a)}\}$. Note that the delay-domain features $\{\mathbf{f}^{(\rm d,m)}_{k}\}$ after the disentanglement are still biased, they are further processed by the delay-domain alignment module. Then, delay-domain alignment in Fig.~\ref{fig: framework}(d) is applied to address the peak bias and the power leakage effect, which fundamentally guarantees the generalizability of the trained DL-extrapolator. 

\section{Implementation of Physics-Based Disentanglement and Alignment Framework}
In this section, the implementation of the physics-based disentanglement and alignment framework is detailed. In Sec.~\ref{subsec: MCD}, the SVD-based multi-cluster decoupling is verified for both near-field and far-field. Then, angle-delay disentanglement is proposed in Sec.~\ref{subsec: disentanglement}, which is supported by real-world measurement data. Next, delay-domain alignment for debiasing is proposed in Sec.~\ref{subsec: delay-domain alignment}. 
\label{sec: implementation}

\subsection{Multi-Cluster Decoupling}
\label{subsec: MCD}
The objective of multi-cluster decoupling is to decompose the original channel into a summation of multiple clusters. To support the unified generalization, the multi-cluster decoupling algorithm should be effective in both near-field and far-field scenarios. In \cite{arxiv_wang_2025_Generalizable}, an SVD-based multi-cluster decoupling is proposed based on the intra-cluster and inter-cluster properties in the far-field scenarios. The intra-cluster property indicates that the power of the cluster response matrix is concentrated in the largest singular value, which can be approximated by a rank-one matrix. The inter-cluster property proposes that different clusters are nearly orthogonal in both the angular domain and the delay domain. Thus, to justify the applicability of SVD for unified multi-cluster decoupling, we need to re-examine the intra-cluster and inter-cluster properties in the near-field scenarios, which is detailed as follows.

To simulate the physics-grounded cluster characteristics in the near-field scenarios, we adopt the geometric-based stochastic channel model in near-field (GBSM-NF) \cite{jsac_2025_far_miao,twc_2022_channel_cui} to generate the cluster responses. To ground the cluster behaviour, we assume $N_{\rm p}$ paths are generated within the non-line-of-sight (NLOS) clusters, where the interacted positions are randomly distributed within a sphere with radius $R_{\rm s}$. Here, we assume that $R_{\rm s}=0.5$ m and $N_{\rm p}=5$. The number of clusters in LOS and NLOS scenarios is set as 5, and the Rician factor in LOS scenarios is set as 5 dB. We consider a half-wavelength ULA BS with 32 antennas, where the carrier frequency is 3.4 GHz. System bandwidth is set as 40 MHz, and the number of subcarriers is 32. Since the Rayleigh distance is $45.2$ m, we assume that the user and scatters are randomly distributed within a 120$^\circ$ sector with a range $5\sim45$m, which falls into the near-field region. For the near-field cluster $\mathbf{H}_{l}^{(\rm m, NF)}=\sum_{i=1}^{N_{{\rm r}, i}}\alpha_{l,i}e^{-{\rm j}2\pi f^{(\rm m)}_{1}\tau_{l,i}}\mathbf{a}^{(\rm NF)}(\mathbf{x}_{l,i})\mathbf{b}^{H}_{x}(\tau_{l,i})$, we adopt the rank-one concentration metric $\gamma=\sigma_{l,1}^{2}/\Vert\mathbf{H}^{(\rm m,NF)}_{l}\Vert_F^2$, where $\sigma_{l,1}$ denotes the largest singular value of $\mathbf{H}^{(\rm m,NF)}_{l}$. Based on Monte-Carlo trials, the averages of $\gamma$ in the LOS and NLOS scenarios are 0.994 and 0.993, respectively. Thus, it indicates that the power of the cluster is concentrated in the largest singular value. The rationale lies in the fact that the cluster power is concentrated in the delay-domain. Thus, although the angular-domain power distribution becomes dispersive due to the spherical wave propagation, the rank-one property of the near-field cluster is still held. Next, the inter-cluster level orthogonality property in near-field is investigated. Here, the normalized row orthogonality $\eta_{\rm r}$ and column orthogonality $\eta_{\rm c}$ for different clusters $\mathbf{H}_{l}^{(\rm m,NF)}$ and $\mathbf{H}_{l^\prime}^{(\rm m,NF)}$ are adopted for evaluation \cite{arxiv_wang_2025_Generalizable}, which are defined as 
\begin{equation}
    \begin{aligned}
    \eta_{\rm r}&=\frac{\Vert\big(\mathbf{H}_{l}^{(\rm m,NF)}\big)^{H}\mathbf{H}_{l^\prime}^{(\rm m,NF)}\Vert_{F}}{\Vert(\mathbf{H}^{(\rm m,NF)})^{H}\mathbf{H}^{(\rm m,NF)}\Vert_{F}},\\
    \eta_{\rm c}&=\frac{\Vert\mathbf{H}_{l}^{(\rm m, NF)}\big(\mathbf{H}_{l^\prime}^{(\rm m, NF)}\big)^{H}\Vert_{F}}{\Vert\mathbf{H}^{(\rm m,NF)}\big(\mathbf{H}^{(\rm m, NF)}\big)^H\Vert_{F}}.
    \end{aligned}
\end{equation}
Under the aforementioned simulation settings, the averages of $\eta_{\rm r}$ and $\eta_{\rm c}$ are presented in Table~\ref{tab: orthogonality}. It can be found that near-field clusters are also approximately orthogonal in the angular and delay domains. Since the rank-one intra-cluster property and the dual-orthogonal inter-cluster property are also held for near-field clusters, the unified SVD-based multi-cluster decoupling \cite{arxiv_wang_2025_Generalizable} can seamlessly adapt to the planar and spherical wavefronts and is applicable in both near-field and far-field scenarios, which is given as follows. 

\begin{table}[t]
  \centering
  \belowrulesep=0.5pt
  \aboverulesep=0pt
  \begin{threeparttable}
  \captionsetup{font=footnotesize}
  \caption{Orthogonality of simulated near-field clusters}
    \label{tab: orthogonality}%
    \begin{tabular}{c|c|c}
    \toprule
    Scenarios          & LOS   & NLOS \\
    \midrule
    Average row orthogonality $\eta_{\rm r}$ & 0.0108 & 0.0221 \\
    \midrule
    Average column orthogonality $\eta_{\rm c}$ & 0.0319 & 0.0586 \\
    \bottomrule
    \end{tabular}%
  \end{threeparttable}
  \vspace{-10pt}
\end{table}%

\begin{theorem}
    \label{theo: svd}
    In both near-field and far-field scenarios, clusters in original channel $\mathbf{H}^{(\rm m)}$ can be decoupled with SVD $\mathbf{H}^{(\rm m)}=\sum_{k=1}^{\text{rank}(\mathbf{H}^{(\rm m)})}\sigma_{k}\mathbf{u}_{k}\big(\mathbf{v}_{k}^{(\rm m)}\big)^H$, where $\sigma_{k}$ denotes the $k$th largest singular value. The $k$th decoupled cluster can be represented by $\mathbf{C}^{(\rm m)}_{k}=\sigma_{k}\mathbf{u}_{k}\big(\mathbf{v}_{k}^{(\rm m)}\big)^H$.
\end{theorem}

\subsection{Angle-Delay Disentanglement}
\label{subsec: disentanglement}
The objective of angle-delay disentanglement is to represent the decoupled cluster $\mathbf{C}^{(\rm m)}_{k}$ with angular-domain feature $\mathbf{f}^{(\rm a)}_{k}$ and delay-domain feature $\mathbf{f}^{(\rm d,m)}_{k}$. Based on the physics meanings discussed in Sec.~\ref{subsec: physics-ddaa}, the angular-domain feature $\mathbf{f}^{(\rm a)}_{k}$ is frequency-independent while the delay-domain feature $\mathbf{f}^{(\rm d,m)}_k$ is frequency-dependent. Through disentanglement, the frequency-domain channel extrapolation is transformed into the extrapolation of the delay-domain features $\{\mathbf{f}_{k}^{(\rm d,m)}\}$. From \textbf{Theorem}~\ref{theo: svd}, the row-space (angular-domain) of the decoupled cluster $\mathbf{C}_{k}^{(\rm m)}$ is spanned by the left singular vector $\mathbf{u}_{k}$, and the column-space (delay-domain) is spanned by the right singular vector $\mathbf{v}_{k}^{(\rm m)}$. Therefore, the angle-delay disentanglement followed by the SVD-based multi-cluster decoupling is given as follows. 
\begin{proposition}
    \label{prop: disentanglement}
    With the SVD-based multi-cluster decoupling $\mathbf{H}^{(\rm m)}=\sum_{k=1}^{\text{rank}(\mathbf{H}^{(\rm m)})}\sigma_{k}\mathbf{u}_k\mathbf{v}_k^H$, the angular-domain features $\{\mathbf{f}^{(\rm a)}_k\}$ and delay-domain features $\{\mathbf{f}^{(\rm d,m)}_k\}$ can be formulated as 
    \begin{equation}
    \label{equ: disentanglement}
    \mathbf{f}^{(\rm a)}_{k}=\mathbf{u}_{k},\quad\mathbf{f}^{(\rm d,m)}_k=\sigma_{k}\cdot\text{conj}\left(\mathbf{v}_{k}^{(\rm m)}\right).
\end{equation}
\end{proposition}
Intutively, the decoupled cluster can be naturally reconstructed by  $\mathbf{C}_{k}^{(\rm m)}=\mathbf{f}_{k}^{(\rm a)}\big(\mathbf{f}_{k}^{(\rm d,m)}\big)^T$, which verifies the invertibility of the disentanglement operation. Next, it can be proved that the disentanglement in \eqref{equ: disentanglement} satisfies the desired physics meanings of angular-domain and delay-domain features, which are detailed as follows. 

\begin{figure}[t]
    \vspace{-15pt}
    \captionsetup{font=footnotesize,justification=centering}
    \subfloat[RENEW dataset (far-field) 
    ]{\includegraphics[width=0.24\textwidth]{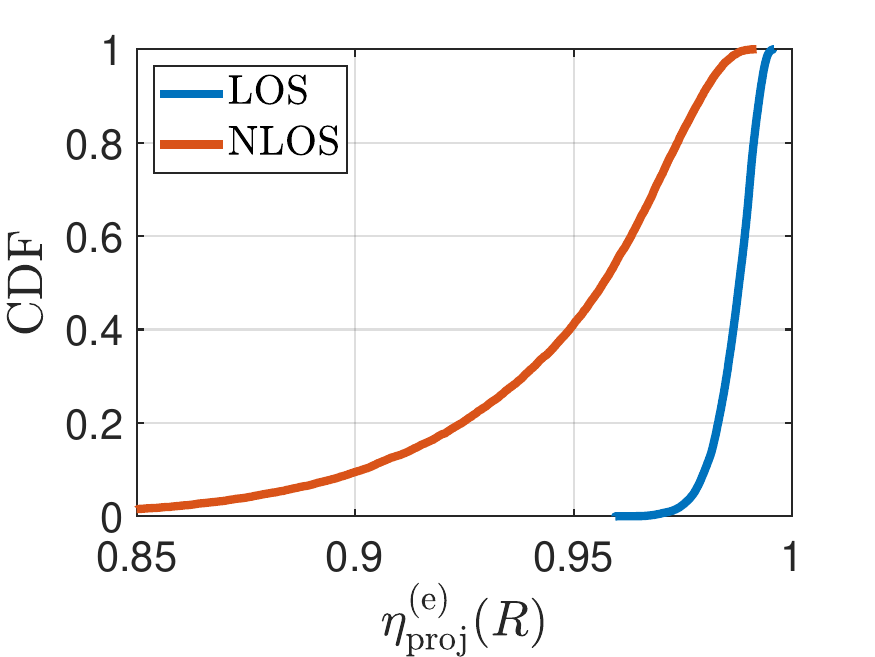}}\captionsetup{font=footnotesize,justification=centering}
    \subfloat[ESPARGOS dataset (near-field)
    ]{\includegraphics[width=0.24\textwidth]{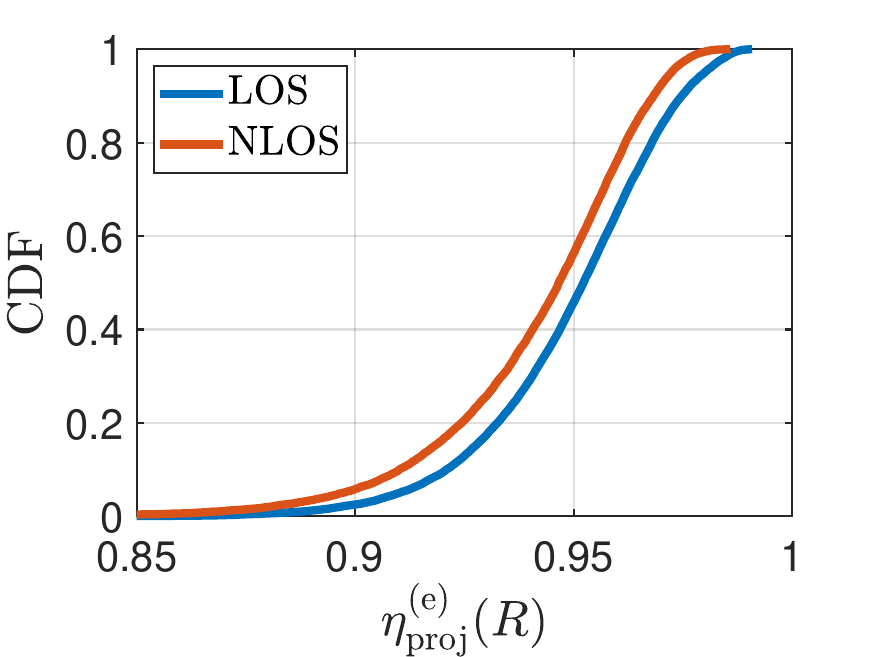}}\captionsetup{font=footnotesize,justification=raggedright}
    \captionsetup{font=small}
    \caption{Validation of frequency-independent angular-domain features based on real-world measurement data.}
    \label{fig: measurement validation}
    \vspace{-10pt}
\end{figure}

The frequency-independence of the angular-domain features can be supported by the real-world far-field and near-field measurement channel data. Here, the far-field RENEW channel dataset \cite{tvt_du_2022_dataset} and the near-field ESPARGOS dataset \cite{dataset-espargos-0002} are adopted. The measurement settings are detailed in Sec.~\ref{subsec: sim2real}. Denote angular-domain feature matrices $\mathbf{F}^{(\rm a)}(r)=\big[\mathbf{f}_{1}^{(\rm a)},\ldots\mathbf{f}_{r}^{(\rm a)}\big]\in\mathbb{C}^{N_{\rm T}\times r}$ with the largest $r$ singular value. Since angular-domain features $\big\{\mathbf{f}^{(\rm a)}_{l}\big\}_{l=1}^{r}$ are mutually orthogonal and have unit norms, we can measure the power portion of $\mathbf{H}^{(\rm e)}$ when projected on the dominant angular-domain features $\big\{\mathbf{f}^{(\rm a)}_{l}\big\}_{l=1}^{r}$. To this end, two quantities
\begin{equation}
    \begin{aligned}
    \eta_{\text{proj}}^{(\rm m)}(r)&=\frac{\Vert(\mathbf{F}^{(\rm a)}(r))^H\mathbf{H}^{(\rm m)}\Vert_{F}^2}{\Vert\mathbf{H}^{(\rm m)}\Vert_{F}^2},\\
    \eta_{\text{proj}}^{(\rm e)}(r)&=\frac{\Vert(\mathbf{F}^{(\rm a)}(r))^H\mathbf{H}^{(\rm e)}\Vert_{F}^2}{\Vert\mathbf{H}^{(\rm e)}\Vert_{F}^2}
    \end{aligned}
\end{equation}
are defined to evaluate the frequency independence of the angular-domain feature. For each channel instance, we consider the metric $\eta_{\rm proj}^{(\rm e)}(R)$ with $R=\min\big\{r:\eta_{\rm proj}^{(\rm m)}(r)>\eta_{\rm th}\big\}$. Intuitively, when $\eta_{\rm proj}^{(\rm e)}(R)$ approaches $\eta_{\rm th}$, it indicates that the dominant angular-domain features are frequency-independent in the target frequency band. By selecting $\eta_{\rm th}=0.999$, the cumulative density functions (CDFs) of $\eta_{\rm proj}^{(\rm e)}(R)$ in the real-world far-field and near-field scenarios are plotted in Fig.~\ref{fig: measurement validation}. A key observation is that the 10th percentile of $\eta_{\rm proj}^{(\rm e)}(R)$ across all scenarios exceeds 0.9. This provides empirical evidence that the angular-domain features, extracted and optimized solely within the measured band, retain a dominant representation power (>95\% on average) in the target frequency band. Thus, it can be concluded that the angular-domain features are frequency-independent in both near-field and far-field scenarios. 

The physics-meaning of the delay-domain feature $\{\mathbf{f}^{(\rm d,m)}_{k}\}$ can be theoretically verified. Based on the property of SVD, the angular-domain features $\{\mathbf{f}_{k}^{(\rm a)}\}$ are mutually orthogonal, i.e., $\big(\mathbf{f}_{k}^{(\rm a)}\big)^H\mathbf{f}_{k^\prime}^{(\rm a)}=0$ is held for $k\neq k^\prime$. Thus, it can be derived that 
\begin{equation}
    \label{equ: delay-domain feature physics}
    \big(\mathbf{f}^{(\rm a)}_k\big)^{H}\mathbf{H}^{(\rm m)}=\sum_{k^\prime=1}^{\text{rank}(\mathbf{H}^{(\rm m)})}\big(\mathbf{f}_{k}^{(\rm a)}\big)^{H}\mathbf{f}_{k^\prime}^{(\rm a)}\big(\mathbf{f}^{(\rm d,m)}_{k^\prime}\big)^T=(\mathbf{f}^{(\rm d,m)}_{k}\big)^T.
\end{equation}
Thus, the delay-domain feature $\mathbf{f}^{(\rm d,m)}_k$ can be derived as $\mathbf{f}^{(\rm d,m)}_k=\big(\mathbf{H}^{(\rm m)}\big)^T\text{conj}\big(\mathbf{f}_{k}^{(\rm a)}\big)$. Based on the channel model in \eqref{equ: Hm far} and \eqref{equ: Hm near}, the delay-domain feature $\mathbf{f}_{k}^{(\rm d,m)}$ can be further reformulated as 
\begin{equation}
    \label{equ: delay-domain feature physics-new}
    \mathbf{f}_{k}^{(\rm d,m)}=\sum_{l=1}^{N_{\rm cl}}\sum_{i=1}^{N_{{\rm r},i}}\beta_{k,l,i}e^{-{\rm j}2\pi f^{(\rm m)}_{i}\tau_{l,i}}\mathbf{b}_{\rm m}(\tau_{l,i}),
\end{equation}
where the complex gain $\beta_{k,l,i}$ is calculated by 
\begin{equation}
    \label{equ: beta}
    \beta_{k,l,i}=\begin{cases}
    \alpha_{l,i}\mathbf{a}^{(\rm FF)}(\varphi_{l,i},\theta_{l,i})^T\text{conj}\big(\mathbf{u}_{k}\big),\quad\text{far-field;}\\
    \alpha_{l,i}\mathbf{a}^{(\rm NF)}(\mathbf{x}_{l,i})^T\text{conj}\big(\mathbf{u}_{k}\big),\quad\text{near-field.}
    \end{cases}
\end{equation}
Here, the delay-domain feature $\mathbf{f}^{(\rm d,m)}_{k}$ can precisely reserve the underlying path delays in the channel, which directly reflects the delay-domain power distribution. Thus, the channel extrapolation can be reformulated as the extrapolation of the delay-domain features $\big\{\mathbf{f}_{k}^{(\rm d, m)}\big\}$, which achieves the goal of angle-delay disentanglement in physics.

\subsection{Delay-Domain Alignment}
\label{subsec: delay-domain alignment}
The objective of delay-domain alignment is to address the distribution shift of decoupled delay-domain features. As illustrated in Fig.~\ref{fig: framework}(c), although the disentangled delay-domain features are sparse, the peak position varies and is environment-dependent. Additionally, the delay-domain power leakage effect exists due to the intra-cluster delay spread and the imperfect delay-domain orthogonality between clusters. To this end, the oversampled discrete Fourier transformation (DFT) codebook is utilized to debias the delay-domain peak position and mitigate the power leakage effect. Here, the $O$-oversampled DFT codebook is adopted, where the $n$-th codeword $\mathbf{w}^{(\rm m)}_{n}\in\mathbb{C}^{K^{(\rm m)}\times1}$ $(0\leq n\leq OK^{(\rm m)}-1)$ is defined as 
\begin{equation}
    \label{equ: codeword m}
    \mathbf{w}^{(\rm m)}_{n}=\left[1,e^{{\rm j}2\pi\frac{ n}{OK^{(\rm m)}}},\ldots,e^{{\rm j}2\pi\frac{n(K^{(\rm m)}-1)}{OK^{(\rm m)}}}\right]^T.
\end{equation}
Consider the alignment of the delay-domain feature $\mathbf{f}^{(\rm d, m)}_{k}$ as an example. Then, the peak position $n^\star_k$ can be calculated by 
\begin{equation}
    \label{equ: peak pos}
    n^\star_k=\mathop{\arg\max}_{n}\left\{\left|\big(\mathbf{w}^{(\rm m)}_{n}\big)^T\mathbf{f}^{(\rm d,m)}_k\right|^2\right\}.
\end{equation}
With the searched peak position $n^\star_k$, the aligned delay-domain feature $\widetilde{\mathbf{f}}^{(\rm d,m)}_k$ can be calculated by 
\begin{equation}
    \label{equ: aligned feature}
    \widetilde{\mathbf{f}}^{(\rm d,m)}_k=\mathbf{w}^{(\rm m)}_{n^\star_k}\odot\mathbf{f}^{(\rm d,m)}_k.
\end{equation}
Based on \eqref{equ: delay-domain feature physics-new}, the aligned delay-domain feature $\widetilde{\mathbf{f}}^{(\rm d,m)}_k$ of the $l$th decoupled cluster can also be represented by 
\begin{equation}
    \label{equ: aligned feature latent}
    \widetilde{\mathbf{f}}^{(\rm d,m)}_k\!=\!\sum_{l=1}^{N_{\rm cl}}\sum_{i=1}^{N_{{\rm r},i}}\beta_{k,l,i}e^{-{\rm j}2\pi f^{(\rm m)}_{1}\tau_{l,i}}\mathbf{b}_{\rm m}\left(\tau_{l,i}\!-\!\frac{n^\star_k}{OK^{(\rm m)}\Delta f}\right).
\end{equation}
It indicates that a constant shift $-\frac{n^\star_k}{OK^{(\rm m)}\Delta f}$ is applied to all underlying paths in the decoupled delay-domain feature $\mathbf{f}^{(\rm d,m)}_k$, which can effectively address the distribution shift. Owing to the preservation of the underlying physics structure in $\mathbf{f}^{(\rm d,m)}_k$, the aligned delay-domain feature $\widetilde{\mathbf{f}}^{(\rm d,m)}_k$ also be parameterized by the underlying path delays $\{\widetilde{\tau}_{k,l,i}\}$, where $\widetilde{\tau}_{k,l,i}=\tau_{l,l}-\frac{n^\star_k}{OK^{(\rm m)}\Delta f}$. The physics preservation facilitates the label transformation to guarantee accurate frequency-domain correlation, which is detailed Sec.~\ref{subsec: training dataset}. 

\section{Workflow of UNiFi-DLE}
Facilitating with the physics-based disentanglement and alignment, the end-to-end workflow of UNiFi-DLE is detailed. Firstly, the preparation of the training dataset is discussed in Sec.~\ref{subsec: training dataset}, which theoretically preserves the frequency-domain correlation. Then, the pipelines of model training and inference are detailed in Sec.~\ref{subsec: training} and \ref{subsec: inference}. 
\subsection{Training Dataset Preparation}
\label{subsec: training dataset}
\begin{figure}[t]
        \centering
        \includegraphics[width=0.45\textwidth]{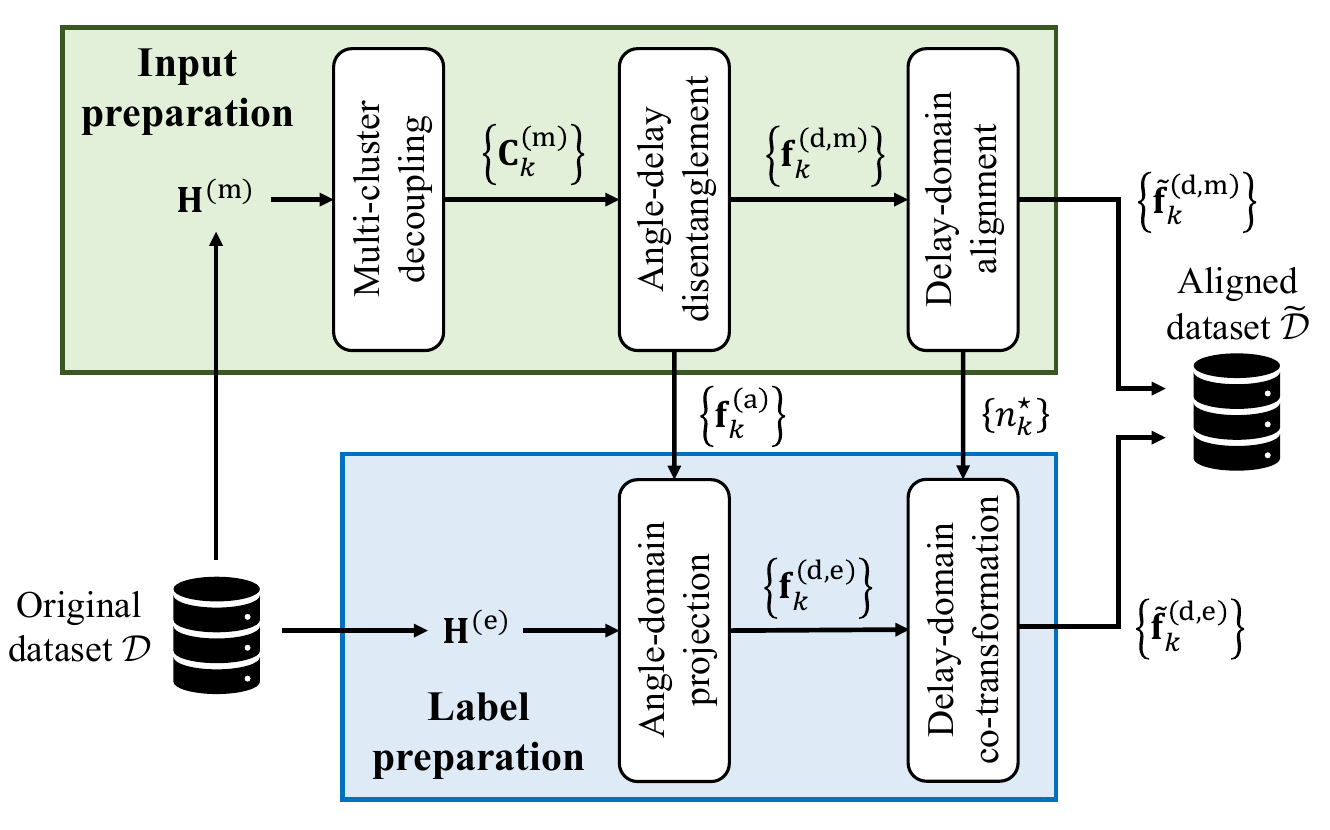}
    \captionsetup{font=footnotesize}
    \caption{Training dataset preparation for UNiFi-DLE.}
    \label{fig: training dataset}
    \vspace{-10pt}
\end{figure}

\begin{figure*}[t]
        \centering
        \includegraphics[width=1\textwidth]{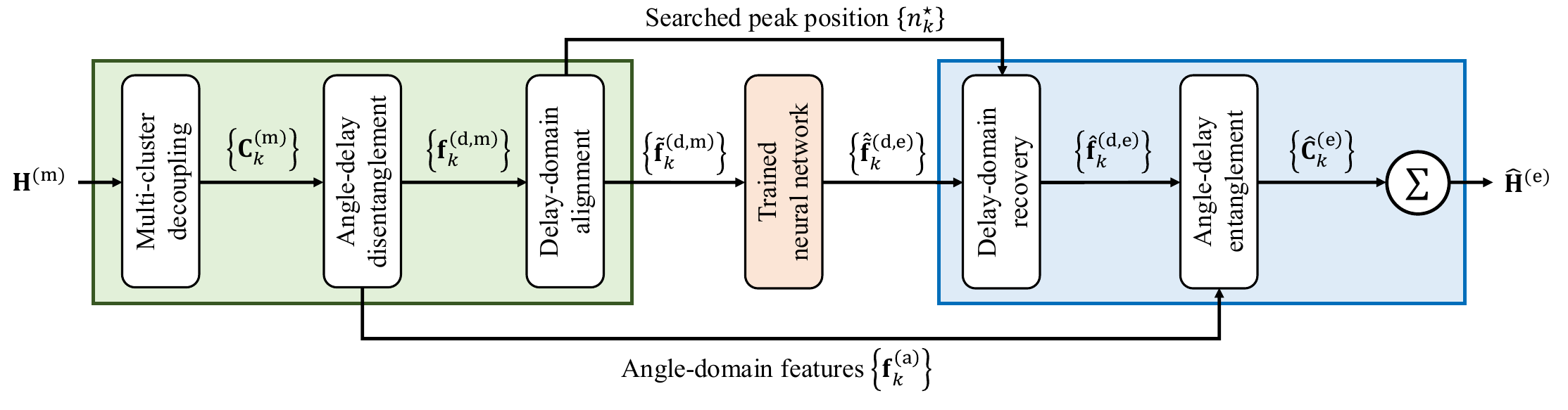}
    \captionsetup{font=footnotesize}
    \caption{End-to-end inference pipeline of UNiFi-DLE, which enables unified generalization for near-field and far-field channel extrapolation.}
    \label{fig: inference}
\end{figure*}

As illustrated in Fig.~\ref{fig: training dataset}, the input and label for the training of the neural network in UNiFi-DLE should be transformed to facilitate the physics-based disentanglement and alignment. For the input preparation, the channel $\mathbf{H}^{(\rm m)}$ at the measured frequency band is firstly sampled from the original channel dataset $\mathcal{D}$. Then, the SVD-based multi-cluster decoupling is applied to $\mathbf{H}^{(\rm m)}$, which yields the decoupled clusters $\{\mathbf{C}^{(\rm m)}_{k}\}$. Based on the angle-delay disentanglement in Sec.~\ref{subsec: disentanglement}, the angular-domain features $\{\mathbf{f}^{(\rm a)}_{k}\}$ and the delay-domain features $\{\mathbf{f}^{(\rm d,m)}_{k}\}$ are obtained from the decoupled clusters $\{\mathbf{C}^{(\rm m)}_{k}\}$. Then, the delay-domain feature $\{\mathbf{f}^{(\rm d,m)}_{k}\}$ is further processed by the delay-domain alignment module in Sec.~\ref{subsec: delay-domain alignment}, which yields the aligned delay-domain features $\{\widetilde{\mathbf{f}}_{k}^{(\rm d,m)}\}$ and the searched peak positions $\{n^{\star}_{k}\}$. Then, the aligned delay-domain features $\{\widetilde{\mathbf{f}}_{k}^{(\rm d,m)}\}$ are served as the neural network input. In UNiFi-DLE, the labels are transformed to preserve the frequency-domain correlation with the measured features $\mathbf{f}_k^{(\rm d,m)}$. This process comprises two steps: angle-domain projection and delay-domain co-transformation. Firstly, the channel $\mathbf{H}^{(\rm e)}$ at the target frequency band is sampled from the original channel dataset $\mathcal{D}$ and input into the angular-domain projection module with the angular-domain features $\{\mathbf{f}^{(\rm a)}_{k}\}$, which yields the delay-domain feature $\{\mathbf{f}^{(\rm d,e)}_k\}$ at the target frequency band with 
\begin{equation}
    \label{equ: delay-feature target}
    \mathbf{f}^{(\rm d,e)}_{k}=\big(\mathbf{H}^{(\rm e)}\big)^T\text{conj}\left(\mathbf{f}_{k}^{(\rm a)}\right). 
\end{equation}
Then, the feature $\{\mathbf{f}^{(\rm d,e)}_k\}$ and the searched peak positions $\{n^\star_k\}$ in the input preparation are further input into the delay-domain co-transformation module to yield the label 

\begin{equation}
    \label{equ: co-transformed label}
    \widetilde{\mathbf{f}}^{(\rm d,e)}_k=e^{{\rm j}\epsilon(n^\star_k)}\big(\mathbf{f}^{(\rm d,e)}_{k}\odot\mathbf{w}^{(\rm e)}_{n^\star_k}\big),
\end{equation}
where 
\begin{equation}
    \label{equ: epsilon}
    \epsilon(n)=2\pi\big(f^{(\rm e)}_1-f^{(\rm m)}_1\big)\frac{n}{OK^{(\rm m)}\Delta f}
\end{equation}
and
\begin{equation}
    \label{equ: we}
    \mathbf{w}^{(\rm e)}_n=\Big[1,e^{{\rm j}2\pi\frac{n}{OK^{(\rm m)}}},\ldots,e^{{\rm j}2\pi\frac{n(K^{(\rm e)}-1)}{OK^{(\rm m)}}}\Big]^T.
\end{equation}
Then, the aligned dataset $\widetilde{\mathcal{D}}$ is yieleded by pairing $\{\widetilde{\mathbf{f}}_{k}^{(\rm d,m)}\}$ and $\{\widetilde{\mathbf{f}}_{k}^{(\rm d,e)}\}$. Compared to the label preparation with parameter estimation and clustering algorithms in our previous work \cite{wang2025generalizable}, the label preparation in UNiFi-DLE can theoretically preserve the frequency-domain correlation, which is given below.

\begin{theorem}
    \label{theo: label correlation}
    The input $\{\widetilde{\mathbf{f}}^{(\rm d,m)}_{k}\}$ and label $\{\widetilde{\mathbf{f}}^{(\rm d,e)}_{k}\}$ is precisely correlated with the phase rotation effect of the underlying path delays.
\end{theorem}
\begin{IEEEproof}
    Denote $\widetilde{\beta}_{k,l,i}=\beta_{k,l,i}e^{{\rm -j}2\pi f^{(\rm m)}_1\frac{n_k^\star}{OK^{(\rm m)}\Delta f}}$. Then, the input $\widetilde{\mathbf{f}}^{(\rm d,m)}_{k}$ the label $\widetilde{\mathbf{f}}^{(\rm d,e)}_{k}$ can be reformulated as
    \begin{equation}
        \label{equ: input and label reformulation}
        \begin{aligned}
        \widetilde{\mathbf{f}}^{(\rm d,m)}_{k}&=\sum_{l=1}^{N_{\rm cl}}\sum_{i=1}^{N_{{\rm r},i}}\widetilde{\beta}_{k,l,i}e^{-{\rm j}2\pi f^{(\rm m)}_{1}\widetilde{\tau}_{k,l,i}}\mathbf{b}_{\rm m}\left(\widetilde{\tau}_{k,l, i}\right),\\
        \widetilde{\mathbf{f}}^{(\rm d,e)}_{k}&=\sum_{l=1}^{N_{\rm cl}}\sum_{i=1}^{N_{{\rm r},i}}\widetilde{\beta}_{k,l,i}e^{-{\rm j}2\pi f^{(\rm e)}_{1}\widetilde{\tau}_{k,l,i}}\mathbf{b}_{\rm e}\left(\widetilde{\tau}_{k,l, i}\right).
        \end{aligned}
    \end{equation}
    It can be found that the input $\widetilde{\mathbf{f}}^{(\rm d,m)}_{k}$ the label $\widetilde{\mathbf{f}}^{(\rm d,e)}_{k}$ shares the common compelx gains $\{\widetilde{\beta}_{k,l,i}\}$ and the delays $\{\widetilde{\tau}_{k,l,i}\}$, and correlates with the phase rotation of the delays $\{\widetilde{\tau}_{k,l,i}\}$.
\end{IEEEproof}
In the dataset preparation stage, the universal frequency-domain correlation between the input and label is guaranteed, which facilitates the unified generalization of UNiFi-DLE in both near-field and far-field scenarios.

\subsection{Training Stage}
\label{subsec: training}
The neural network in UNiFi-DLE is trained with the aligned dataset $\widetilde{\mathcal{D}}$ to guarantee the model generalizability in both near-field and far-field scenarios. Explicitly, the neural network $\xi$ in UNiFi-DLE is optimized with the loss function 
\begin{equation}
    \label{equ: loss function}
    \mathcal{J}=\mathbb{E}\left\{\left\|\widetilde{\mathbf{f}}^{(\rm d,e)}_{k}-\xi\left(\widetilde{\mathbf{f}}^{(\rm d,m)}_{k}\right)\right\|_{2}^{2}\right\}.
\end{equation}
Then, the neural network $\xi$ in UNiFi-DLE can focus on the frequency-domain correlation of the delay-domain features and avoid fitting heterogeneous wave propagation in the near-field and far-field scenarios. Thus, UNiFi-DLE can be trained in a unified manner with both near-field and far-field channel data, which enhances the model generalizability in practical deployment. 
\begin{remark}
    \label{remk: array size}
    The UNiFi-DLE can also generalize to different array shapes. Explicitly, the input and output dimensions of the neural network in UNiFi-DLE are $K^{(\rm m)}$ and $K^{(\rm e)}$, which are independent of the antenna number $N_{\rm T}$. Thus, UNiFi-DLE can generalize to different array shapes with the same frequency band settings.
\end{remark}
\subsection{Inference Stage}
\label{subsec: inference}
The objective of model inference is to accurately extrapolate the full channel matrix to the target frequency band, where the end-to-end pipeline is illustrated in Fig.~\ref{fig: inference}. Firstly, the physics-based disentanglement and alignment modules in Sec.~\ref{sec: implementation} are applied to the input channel $\mathbf{H}^{(\rm m)}$ at the measured frequency band, which yields the angular-domain feature $\{\mathbf{f}_{k}^{(\rm a)}\}$, the searched peak positions $\{n^\star_k\}$ and the aligned delay-domain features $\{\widetilde{\mathbf{f}}^{(\rm d,m)}_{k}\}$. Then, the aligned delay-domain features $\{\widetilde{\mathbf{f}}^{(\rm d,m)}_{k}\}$ are input into the trained neural network in UNiFi-DLE, which yields the extrapolated aligned delay-domain features at the target frequency band $\{\widehat{\widetilde{\mathbf{f}}}^{(\rm d,e)}_{k}\}$. Then, the searched peak position $\{n^{\star}_k\}$ and $\{\widehat{\widetilde{\mathbf{f}}}^{(\rm d,e)}_{k}\}$ are input into the delay-domain recovery module to recover the delay bias and the phase rotation term, which yields 
\begin{equation}
    \label{equ: delay-domain recovery}
    \widehat{\mathbf{f}}^{(\rm d,e)}_k=e^{-{\rm j}\epsilon(n_k^\star)}\left(\widehat{\widetilde{\mathbf{f}}}_{k}^{(\rm d,e)}\odot\text{conj}\big(\mathbf{w}^{(\rm e)}_{n_k^\star}\big)\right).
\end{equation}
Next, the recovered delay-domain features $\{\widehat{\mathbf{f}}_{k}^{(\rm d,e)}\}$ and the angular-domain features $\{\mathbf{f}_{k}^{(\rm a)}\}$ are input into the angle-delay entanglement module, which yields the extrapolated response $\widehat{\mathbf{C}}^{(\rm e)}_k$ of the decoupled cluster with 
\begin{equation}
    \label{equ: extrapolated cluster}
    \widehat{\mathbf{C}}^{(\rm e)}_k=\mathbf{f}^{(\rm a)}_k(\widehat{\mathbf{f}}^{(\rm d,e)}_k)^T.
\end{equation}
Thus, the extrapolated channel $\widehat{\mathbf{H}}^{(\rm e)}$ can be obtained by summing up all the extrapolated clusters, i.e., $\widehat{\mathbf{H}}^{(\rm e)}=\sum_{k}\widehat{\mathbf{C}}^{(\rm e)}_k$. 
\begin{remark}
    \label{remk: universality}
    The proposed UNiFi-DLE serves as a versatile, generalizable learning framework capable of accommodating diverse neural network architectures. This is grounded in the principle that the input delay-domain features can be effectively aligned between the training and test environments. Meanwhile, based on \textbf{Theorem}~\ref{theo: label correlation}, the universal frequency-domain correlation is precisely preserved in the training dataset preparation. Since the neural network exhibits universal approximation capability to fit complex correlations, the UNiFi-DLE is compatible with various existing neural network structures. 
\end{remark}

\section{Experiment Results}
In Sec.~\ref{subsec: setup}, the simulation setups are detailed. Then, extensive simulations are conducted in Sec.~\ref{subsec: simulation}, which validate the unified generalizability of the proposed UNiFi-DLE. Further, sim-to-real results are provided to prove the effectiveness of the proposed UNiFi-DLE in real-world scenarios. 
\subsection{Simulation Setup}
\label{subsec: setup}
\begin{table*}[htbp]
  \centering
  \belowrulesep=0pt
  \aboverulesep=0pt
  \captionsetup{font=footnotesize}
  \caption{Unified Generalizability Comparison of Near-Field and Far-Field Channel Extrapolation Over Simulated Datasets in dB}
    \begin{tabular}{c|c|c|c|c|c|c|c|c}
    \toprule
    \multicolumn{1}{c|}{\multirow{2}{*}{\centering Test  dataset}} & \multicolumn{1}{c|}{\multirow{2}{*}{\centering Training dataset}} & \multirow{2}{*}{\centering Vanilla-DLE \cite{Asilomar_Alrabeiah_2019_deep}} & \multicolumn{3}{c|}{Vanilla-DLE w/ DA \cite{twc_liu_2024_deep,chinacom_han_2024_AI,jstsp_guo_2022_user}} & \multirow{2}{*}{\centering HORCRUX \cite{mobicom_Banerjee_2024_HORCRUX}} & \multirow{2}{*}{\centering PO-DLE+PA \cite{wang2025generalizable}} & \multirow{2}{*}{\centering UNiFi-DLE} \\
\cmidrule{4-6}          &       &       & ADS   & flipping  & RPS    &       &       &  \\
    \midrule
    \multirow{3}{*}{\centering WAIR-D} & UMa   & 0.84  & 0.78  & 0.86  & 0.13  & \multirow{3}{*}{\centering -0.09} & -10.92 & \textbf{-11.47} \\
\cmidrule{2-6}\cmidrule{8-9}          & GBSM-NF-Tr & 2.26  & 2.71  & 1.98  & 2.20   &       & -8.98 & \textbf{-12.99} \\
\cmidrule{2-6}\cmidrule{8-9}          & UMa+GBSM-NF-Tr & 0.35  & 0.42  & 0.52  & -0.15 &       & -11.60 & \textbf{-12.99} \\
    \midrule
    \multirow{3}{*}{\centering GBSM-NF-Te} & UMa   & 1.72  & 1.61  & 1.23   & 1.34  & \multirow{3}{*}{\centering -9.64} & -6.63 & \textbf{-11.94} \\
\cmidrule{2-6}\cmidrule{8-9}          & GBSM-NF-Tr & 3.16  & 3.09  & 3.06  & 3.12  &       & -12.02 & \textbf{-15.99} \\
\cmidrule{2-6}\cmidrule{8-9}          & UMa+GBSM-NF-Tr & 1.71  & 1.75  & 1.56  & 1.59  &       & -10.70 & \textbf{-15.97} \\
    \bottomrule
    \end{tabular}%
  \label{tab: unified generalizability}%
\end{table*}%

In the simulations, we consider three types of simulated far-field and near-field channel datasets, and the default settings are detailed as follows.
\begin{enumerate}
    \item \textbf{UMa} \cite{3gpp.38.901}: This dataset is generated based on the urban macro (UMa) scenario in 3GPP TR 38.901 document. Here, the LOS channel is composed of 12 clusters, and the NLOS channel is composed of 20 clusters, where 20 paths are generated within a cluster. The BS is equipped with a UPA with 8 horizontal antennas and 4 vertical antennas. The starting frequencies $f^{(\rm m)}_1$ and $f^{(\rm e)}_1$ are set as 3.5 GHz and 3.51 GHz, the numbers of subcarriers are set as $K^{(\rm m)}=K^{(\rm e)}=32$, and the spacing is set as $\Delta f=0.3125$ MHz. Since the Rayleigh distance is 3.42 m, and the horizontal user distance ranges from 35 m to 250 m, this dataset falls in the far-field scenario. This dataset is used for model training and contains $10^5$ samples. 
    \item \textbf{WAIR-D} \cite{huangfu2022wair}: This dataset is generated from 100 real-world city maps based on a precise ray-tracer. The BS is equipped with a UPA with 8 horizontal antennas and 4 vertical antennas. The starting frequencies $f^{(\rm m)}_1$ and $f^{(\rm e)}_1$ are set as 2.6 GHz and 2.61 GHz, the numbers of subcarriers are set as $K^{(\rm m)}=K^{(\rm e)}=32$, and the spacing is set as $\Delta f=0.3125$ MHz. The Rayleigh distance is 4.62 m and falls in a far-field scenario. The dataset size has $10^4$ samples in total and is used for test. 
    \item \textbf{GBSM-NF}: This dataset falls in the near-field scenario. The generation process and default cluster settings have been discussed in Sec.~\ref{subsec: MCD}. For the training dataset, an NLOS scenario with 5 clusters is considered, where the distances from the BS to the user and scatters are selected from $U(5{\rm m},45{\rm m})$. The starting frequencies $f^{(\rm m)}_1$ and $f^{(\rm e)}_1$ are set as 3.4 GHz and 3.44 GHz, and the spacing is set as $\Delta f=1.25$ MHz. For the test dataset, a mixed LOS-NLOS scenario with LOS probability $P_{\rm LOS}=0.5$ is considered. The number of clusters in both LOS and NLOS status is randomly chosen from $U(1,5)$, and the Rician factor for LOS status is set as 5 dB. The distances from BS to the user and scatters are randomly selected from $U(5\text{m}, 25\text{m})$. The starting frequencies $f^{(\rm m)}_1$ and $f^{(\rm e)}_1$ are set as 3.4 GHz and 3.41 GHz, and the spacing is set as $\Delta f=0.3125$ MHz. The number of subcarriers in both training and test datasets is set as the number of subcarriers are set as $K^{(\rm m)}=K^{(\rm e)}=32$. The training dataset size is set as $10^5$, and the test dataset size is set as $10^4$. For expression clarity, we refer to the training and test datasets as GBSM-NF-Tr and GBSM-NF-Te, respectively. 
\end{enumerate} 
Further, we consider the following baselines for comparison:
\begin{figure}[t]
    \vspace{-10pt}
        \centering
        \includegraphics[width=0.5\textwidth]{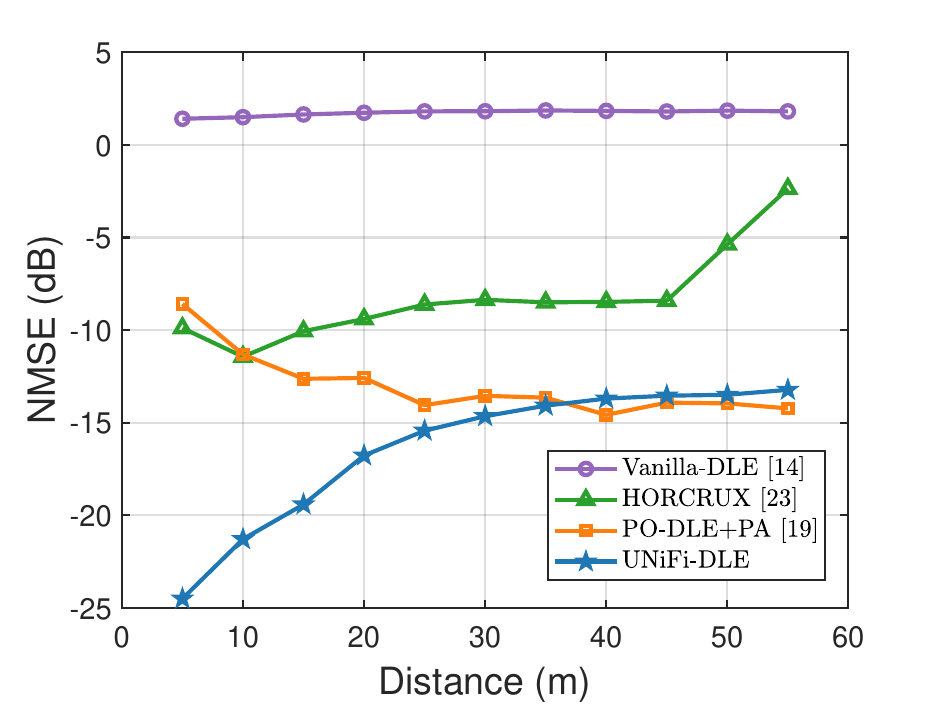}
    \captionsetup{font=footnotesize}
    \caption{Generalizability comparison under different distances in the dataset GBSM-NF-Te. }
    \label{fig: generalizability distance}
    \vspace{-10pt}
\end{figure}
\begin{enumerate}
    \item \textbf{Vanilla-DLE} \cite{Asilomar_Alrabeiah_2019_deep}: This baseline directly inputs the measured channel $\mathbf{H}^{(\rm m)}$ into the neural network to approximte $\mathbf{H}^{(\rm e)}$ at the target frequency band. 
    \item \textbf{Vanilla-DLE w/ DA} \cite{twc_liu_2024_deep,chinacom_han_2024_AI,jstsp_guo_2022_user}: This baseline applies data augmentation (DA) techniques to the training dataset of Vanilla-DLE. Specific DA techniques include ADS \cite{twc_liu_2024_deep}, flipping \cite{chinacom_han_2024_AI}, and RPS \cite{jstsp_guo_2022_user}. 
    \item \textbf{HORCRUX} \cite{mobicom_Banerjee_2024_HORCRUX}: This baseline is composed of eight parallel neural network channel dividers and eight mini-neural network distance estimators, where the channel is extrapolated in an antenna-wise manner. The training details and neural network configurations are aligned with \cite{mobicom_Banerjee_2024_HORCRUX}. 
    \item \textbf{PO-DLE+PA} \cite{wang2025generalizable}: This baseline is our previous work. It composes path-oriented DL extrapolator design and path alignment to address the distribution shift for far-field large-scale MIMO channels.
\end{enumerate}
In the proposed UNiFi-DLE and the baselines of Vanilla-DLE, Vanilla-DLE w/ DA, and PO-DLE+PA, the multi-layer perceptron with three hidden layers is adopted by default, where the hidden dimensions are all set as 512. The oversampling factors in UNiFi-DLE and PO-DLE+PA are set to 2. The normalized mean square error (NMSE) is defined as 
\begin{equation}
    \text{NMSE}=\mathbb{E}\left\{\frac{\Vert\widehat{\mathbf{H}}^{(\rm e)}-\mathbf{H}^{(\rm e)}\Vert_F^2}{\Vert\mathbf{H}^{(\rm e)}\Vert_F^2}\right\}
\end{equation}
is adopted as the performance metric. 

\subsection{Simulation Results}
\label{subsec: simulation}
\subsubsection{Unified Generalizability Comparison}
To justify the unified generalizability across near-field and far-field scenarios, we test the proposed UNiFi-DLE in the simulated WAIR-D and GBSM-NF-Te datasets, which are presented in Table~\ref{tab: unified generalizability}. With different training datasets, proposed UNiFi-DLE can robustly achieve the best performance in both WAIR-D and GBSM-NF-Te, where the NMSE has been reduced for 1.39$\sim$5.27 dB compared to the state-of-the-art. Specifically, when trained with the mixed far-field and near-field datasets, the proposed UNiFi-DLE can achieve the NMSE of -12.99 dB and -15.97 dB in WAIR-D and GBSM-NF-Te, which proves that the physics-based disentanglement and alignment can effectively address the heterogeneity across near-field and far-field scenarios. On the contrary, the Vanilla-DLE fails in both OOD near-field and far-field channels, which remains unresolved even with data augmentation techniques. The HORCRUX baseline can reach -9.64 dB NMSE in the GBSM-NF-Te but fails in WAIR-D. Further, the PO-DLE+PA exhibits weaker generalizability compared to the proposed UNiFi-DLE, especially in the near-field dataset. Thus, the motivation and the superiority of the proposed UNiFi-DLE are validated. In the following experiments, we consider the results with the mixed training dataset of UMa and GBSM-NF-Tr for simplicity. 

\subsubsection{Generalizabiltiy with Distance}
To justify the mitigation of distance-dependency, we further investigate the model generalizability under different distances. Explicitly, we vary the distance between the BS to the scatters and user in the GBSM-NF-Te from 5 m to 55 m. Then, the model generalizability of the trained UNiFi-DLE and the baselines are plotted in Fig.~\ref{fig: generalizability distance}. When the distance is decreased from 55 m to 5 m, i.e., from far-field to near-field, the extrapolation NMSE of the proposed UNiFi-DLE can be gradually decreased from -13.5 dB to -24.5 dB. The rationale lies in the fact that the propagation delays of the LOS and NLOS paths in the channel decrease with distance. Note that the rate of phase variation in the response of each path is proportional to the propagation delay. Shorter propagation delays result in less rapid channel variations in the frequency domain and facilitate the alignment of the delay-domain feature. On the contrary, we find that the generalization NMSE of the PO-DLE+PA gradually increases from -13.9 dB to -8.6 dB as the distance decreases. The rationale lies in the dispersive angular profile in the near-field region, which limits the path alignment performance. Thus, the results precisely align with our analysis in Sec.~\ref{subsec: challenge} and \ref{subsec: key insight}, which lends interpretability to the proposed UNiFi-DLE. Meanwhile, we find that the proposed UNiFi-DLE can also converge to the extrapolation performance of PO-DLE+PA as the distance increases, which also achieves accurate channel extrapolation in the far-field region and realizes unified generalization. 
\begin{figure}[t]
    \vspace{-10pt}
    \captionsetup{font=footnotesize,justification=centering}
    \centering
    \subfloat[WAIR-D (far-field) 
    ]{\includegraphics[width=0.24\textwidth]{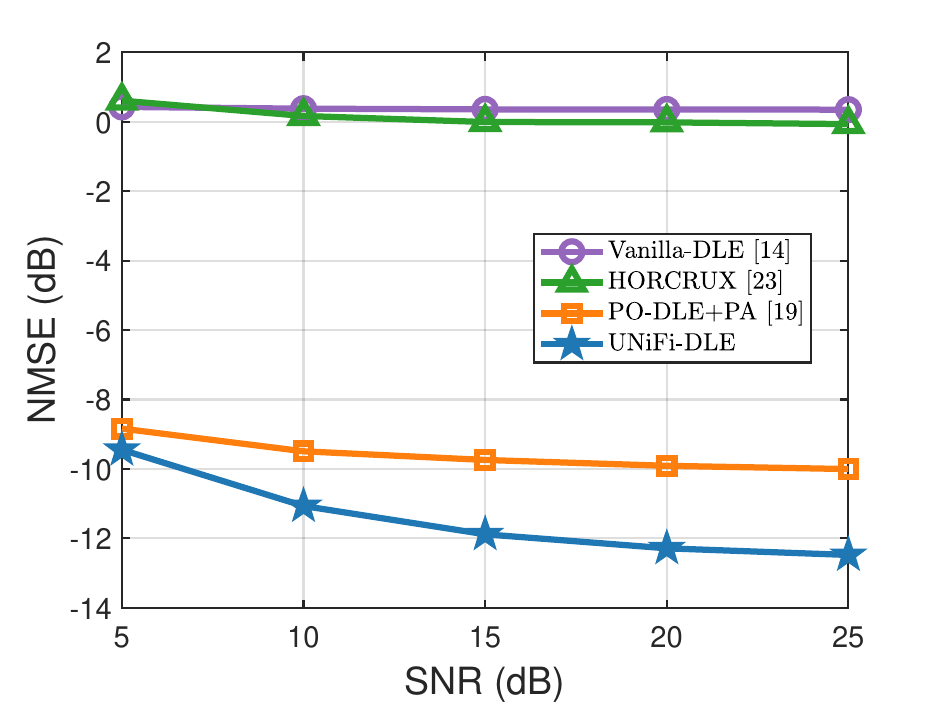}}\captionsetup{font=footnotesize,justification=centering}
    \subfloat[GBSM-NF-Te (near-field) 
    ]{\includegraphics[width=0.24\textwidth]{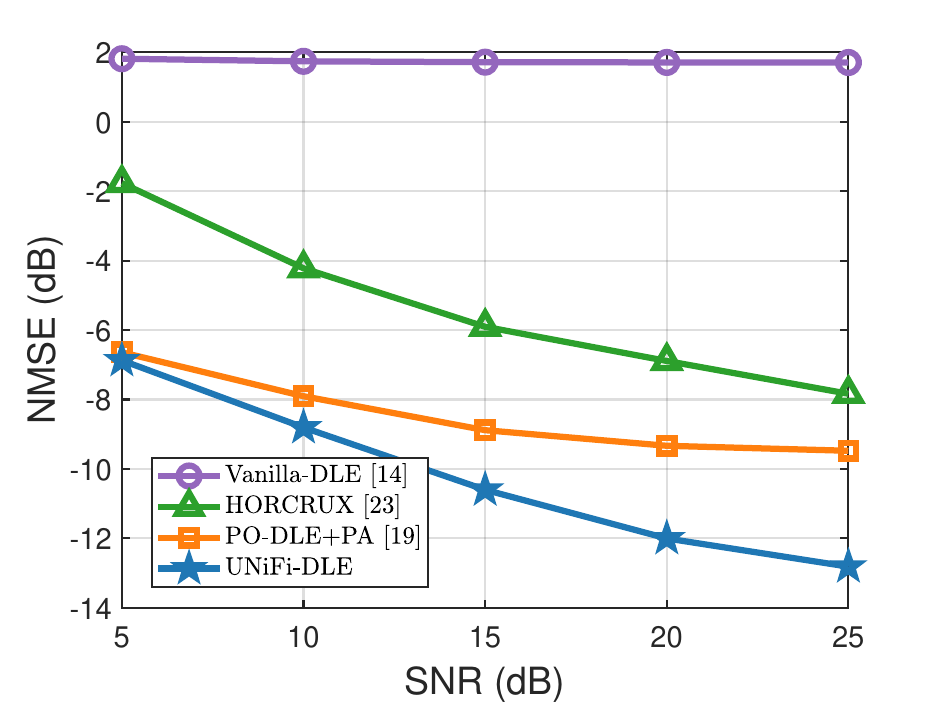}}\captionsetup{font=footnotesize,justification=raggedright}
    \captionsetup{font=small}
    \caption{Generalizability comparison under different SNR.}
    \label{fig: noise robustness}
    \vspace{-10pt}
\end{figure}

\begin{figure}[t]
    \vspace{-10pt}
    \captionsetup{font=footnotesize,justification=centering}
    \centering
    \subfloat[WAIR-D (far-field) 
    ]{\includegraphics[width=0.24\textwidth]{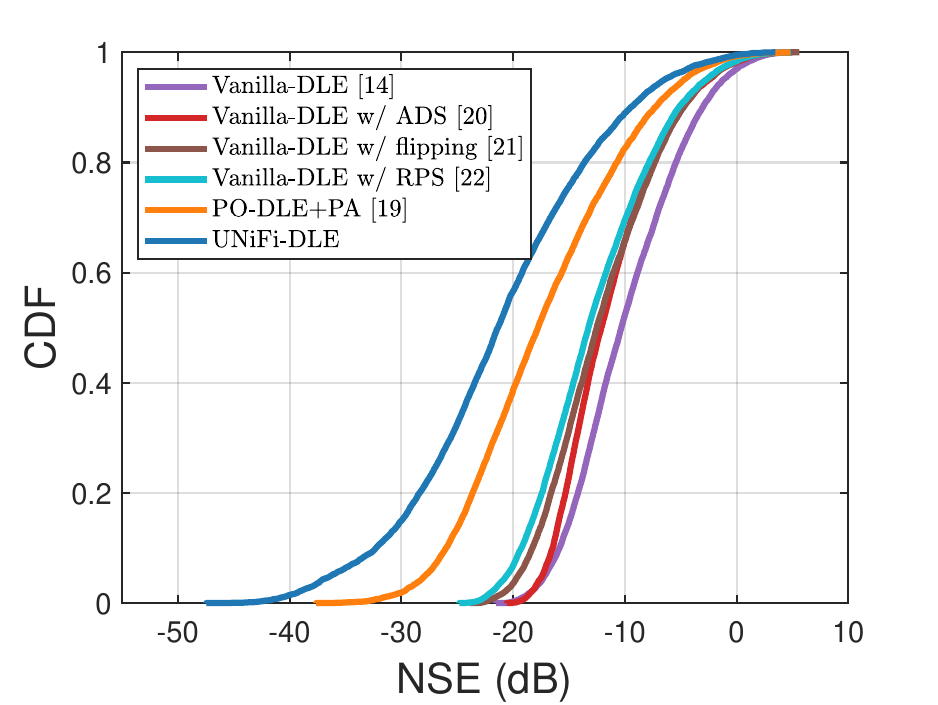}}\captionsetup{font=footnotesize,justification=centering}
    \subfloat[GBSM-NF-Te (near-field) 
    ]{\includegraphics[width=0.24\textwidth]{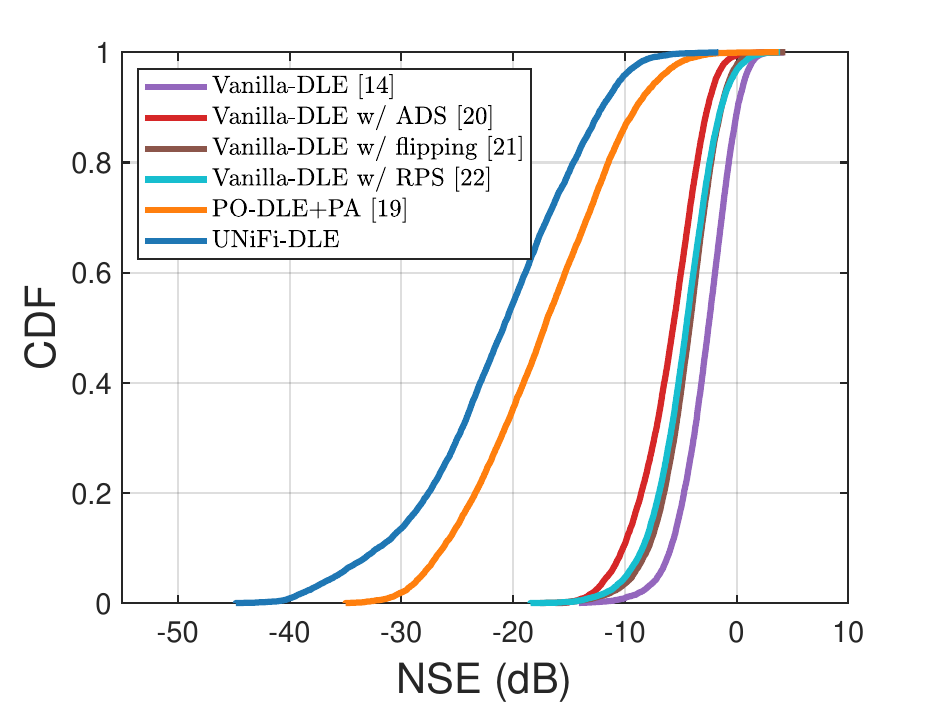}}\captionsetup{font=footnotesize,justification=raggedright}
    \captionsetup{font=small}
    \caption{Unified generalizability comparison, where the sequential LSTM is adopted as the neural network structure for UNiFi-DLE and the baselines.}
    \label{fig: NN structure}
    \vspace{-10pt}
\end{figure}

\subsubsection{Noise Robustness}
Further, the noise robustness of the proposed UNiFi-DLE and the baselines is investigated. Here, we generate the channel samples at the measured frequency band in the training datasets under additive complex Gaussian noise with zero mean, where the signal-to-noise ratio (SNR) of each training sample is selected from $U(5\text{dB}, 25\text{dB})$. Next, the proposed UNiFi-DLE and the baselines are input with the noised channel at the measured frequency band, where the clean channel samples at the target frequency band are served as the supervision signals. After model training, the generalization NMSE of the proposed UNiFi-DLE and the baselines under different SNR is plotted in Fig.~\ref{fig: noise robustness}. It can be observed that the proposed UNiFi-DLE can achieve low NMSE in both WAIR-D and GBSM-NF-Te. The rationale lies in the fact that the SVD-based multi-cluster can capture the correlation of different antennas, which exhibits powerful denoising capability \cite{tit_optshrink_2014_Nadakuditi}. Thus, the proposed UNiFi-DLE is robust to the measurement noise to achieve unified generalization. 

\subsubsection{Neural Network Structure Compatibility}

Next, the unified generalizability of the proposed UNiFi-DLE under different neural network structures is investigated. Here, we adopt the sequential long-short term memory (LSTM) \cite{openj_jiang_2020_deep,wcl_Yao_2024_loss} as the neural network structure in proposed UNiFi-DLE and the baselines of Vanilla-DLE and PO-DLE+PA to model the frequency-domain correlation. Then, the CDF of the normalized square error $\text{NSE}=\Vert\widehat{\mathbf{H}}^{(\rm e)}-\mathbf{H}^{(\rm e)}\Vert_F^2/\Vert\mathbf{H}^{(\rm e)}\Vert_F^2$ of each channel sample are plotted in Fig.~\ref{fig: NN structure} \footnote{Note that the neural networks in HORCRUX do not directly model the frequency-domain correlation. In fact, the channel at the target frequency band in the HORCRUX baseline is calculated based on optimized path parameters. Owing to the distinct objective, we omit the results of HORCRUX under the different neural network structure.}. 
It can be observed that the proposed UNiFi-DLE can robustly achieve the best generalizability in both near-field and far-field scenarios with different neural network structures. Thus, the proposed UNiFi-DLE can serve as a versatile framework for unified generalization independent of specific NN structures. 

\subsubsection{Performance with Array Size}
\begin{figure}[t]
    \vspace{-10pt}
    \captionsetup{font=footnotesize,justification=centering}
    \centering
    \subfloat[WAIR-D (far-field) 
    ]{\includegraphics[width=0.24\textwidth]{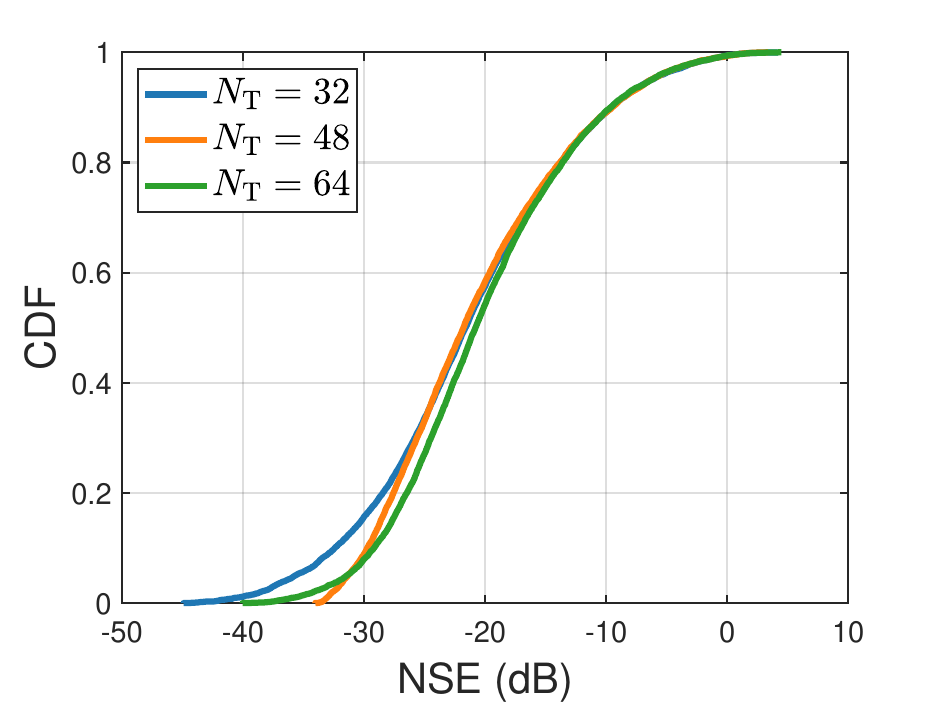}}\captionsetup{font=footnotesize,justification=centering}
    \subfloat[GBSM-NF-Te (near-field) 
    ]{\includegraphics[width=0.24\textwidth]{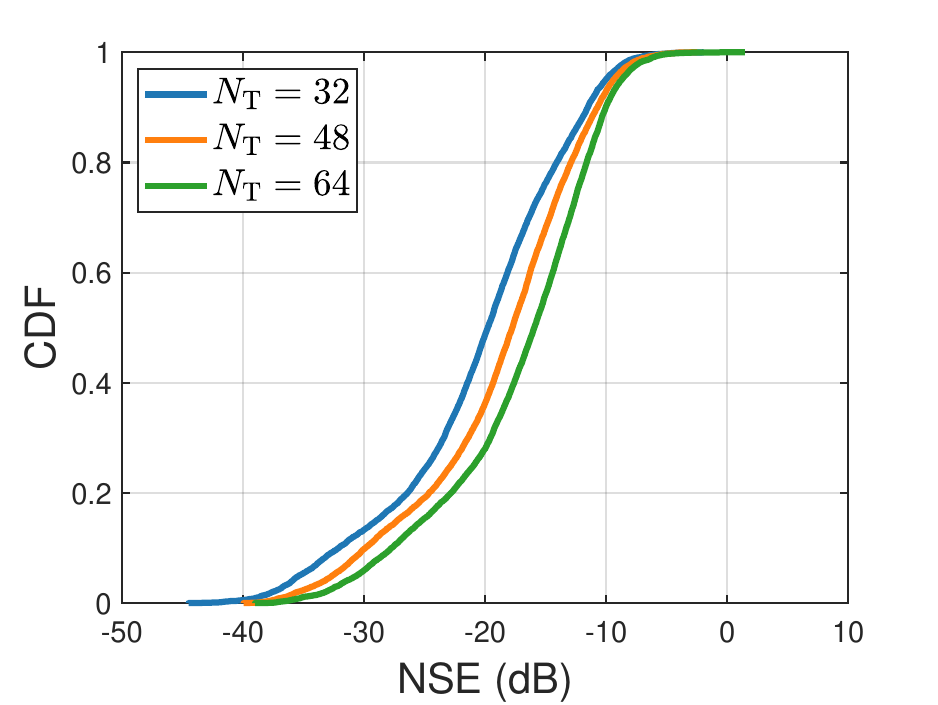}}\captionsetup{font=footnotesize,justification=raggedright}
    \captionsetup{font=small}
    \caption{Generalizability of UNiFi-DLE with different array sizes.}
    \label{fig: arraysize}
    \vspace{-10pt}
\end{figure}

\begin{table}[t]
  \centering
  \belowrulesep=0.1pt
  \aboverulesep=0.1pt
  \captionsetup{font=footnotesize}
  \caption{End-to-end runtime increase comparison in ms}
    \begin{tabular}{c|c|c|c}
    \toprule
    Model & $N_{\rm T}=32$ & $N_{\rm T}=48$ & $N_{\rm T}=64$ \\
    \midrule
    PO-DLE+PA \cite{wang2025generalizable} &    158.75   &   159.15    & 158.89 \\
    \midrule
    UNiFi-DLE & 0.91 & 0.96 & 1.09 \\
    \bottomrule
    \end{tabular}%
  \label{tab: runtime}%
  \vspace{-10pt}
\end{table}%

\begin{figure*}[t]
    \captionsetup{font=footnotesize,justification=centering}
    \centering
    \subfloat[RENEW LOS (far-field) 
    ]{\includegraphics[width=0.24\textwidth]{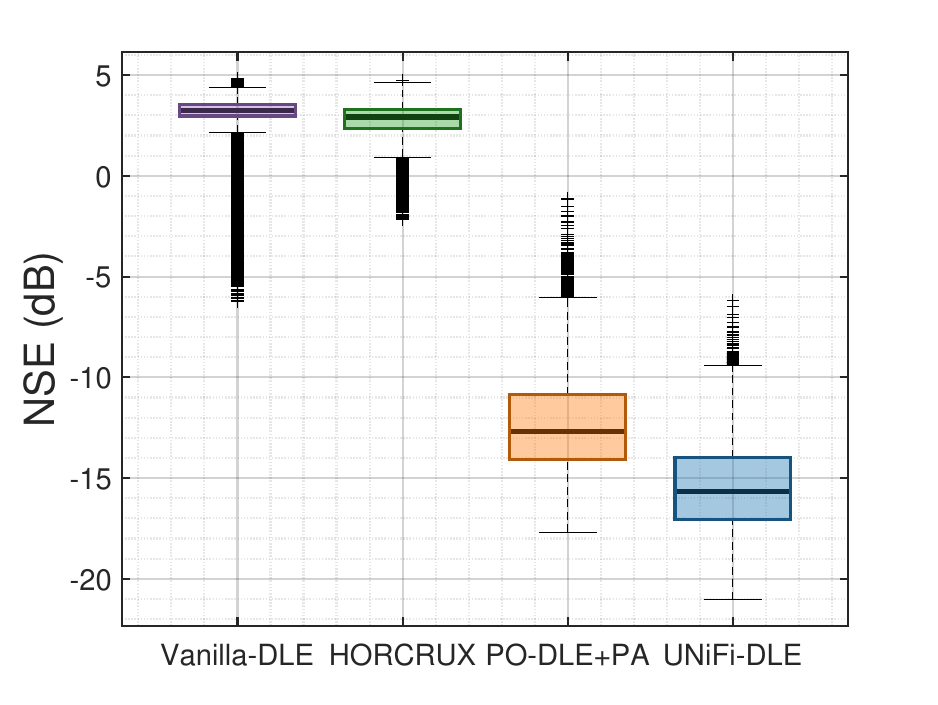}}\captionsetup{font=footnotesize,justification=centering}
    \subfloat[RENEW NLOS (far-field) 
    ]{\includegraphics[width=0.24\textwidth]{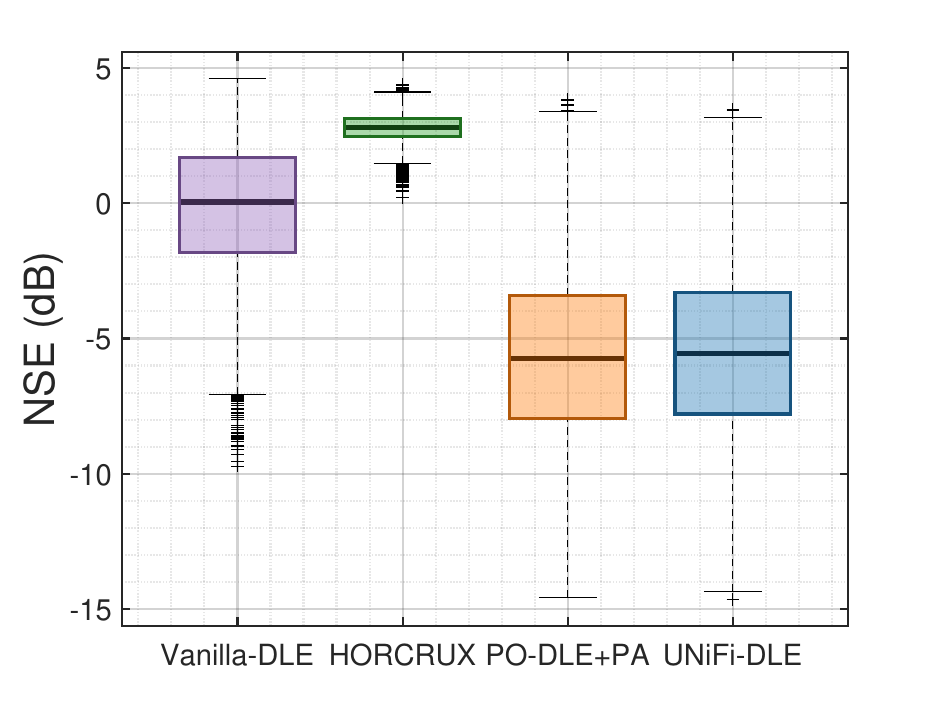}}\captionsetup{font=footnotesize,justification=raggedright}
    \subfloat[ESPARGOS LOS (near-field) 
    ]{\includegraphics[width=0.24\textwidth]{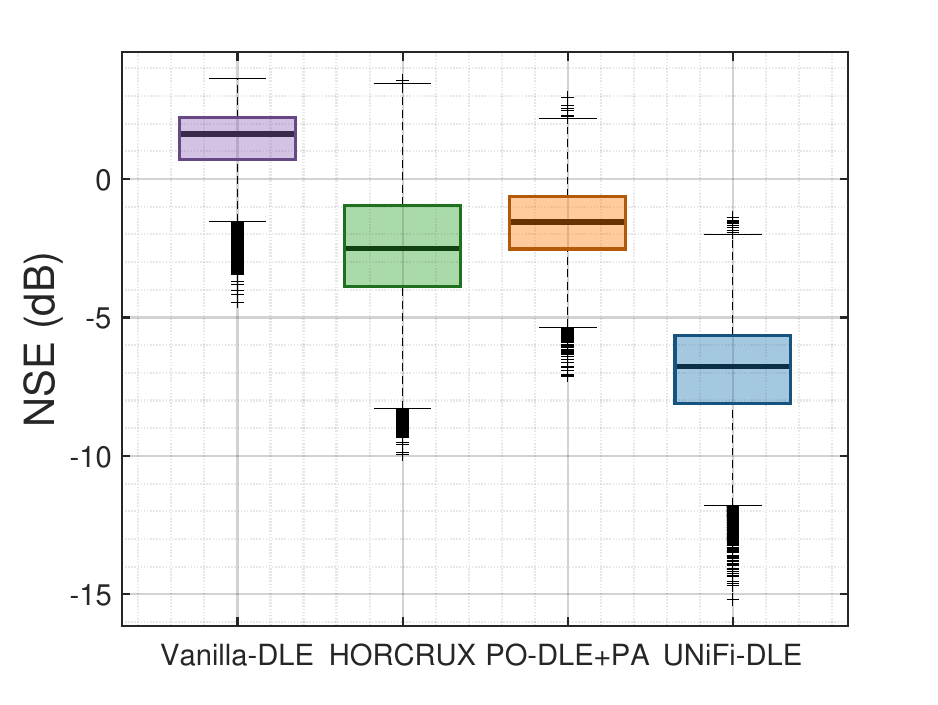}}\captionsetup{font=footnotesize,justification=centering}
    \subfloat[ESPARGOS NLOS (near-field) 
    ]{\includegraphics[width=0.24\textwidth]{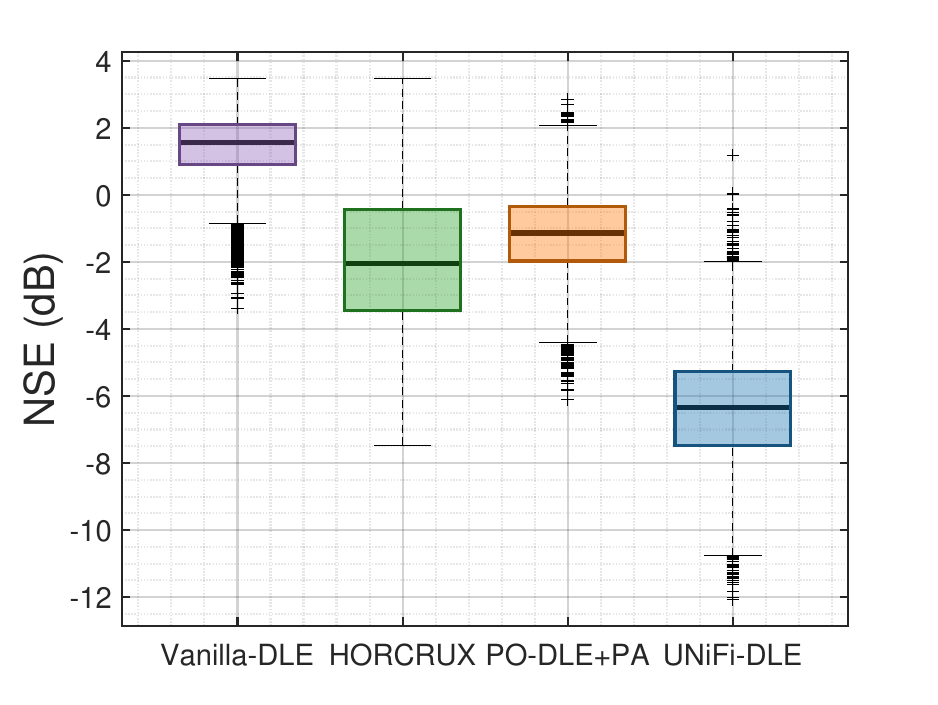}}\captionsetup{font=footnotesize,justification=raggedright}
    \captionsetup{font=small}
    \caption{Sim-to-real generalizability comparison of proposed UNiFi-DLE and the baselines. }
    \label{fig: sim2real}
    \vspace{-5pt}
\end{figure*}
Next, the generalizability of the proposed UNiFi-DLE to different array sizes is further investigated. Explicitly, we fix the number of antennas in the UMa and GBSM-NF-Tr datasets as 32 and vary the number of antennas in the WAIR-D and GBSM-NF-Te datasets from 32 to 64. Then, the CDFs of generalization NSE in WAIR-D and GBSM-NF-Te are plotted in Fig.~\ref{fig: arraysize}. It can be found that the proposed UNiFi-DLE can also accurately extrapolate the channel in larger antenna arrays. Specifically, when the number of antennas increases to 64 in GBSM-NF-Te, the Rayleigh distance increases to 180 m. Then, the trained UNiFi-DLE can still achieve a generalization NMSE with -13.79 dB. The rationale lies in the fact that proposed UNiFi-DLE learns the distributional stable delay-domain features and directly reuses the angular-domain features in channel extrapolation, which verifies \textbf{Remark}~\ref{remk: array size}. Further, we consider the end-to-end runtime of the proposed UNiFi-DLE, which is measured on the Nvidia GeForce RTX 3090 device. Based on the end-to-end pipeline in Fig.~\ref{fig: inference}, the inference time of UNiFi-DLE includes the inference time of the trained neural network and the other processing modules. Thus, to provide a meaningful runtime evaluation compatible with different neural network structures, we consider the end-to-end runtime increases compared to the Vanilla-DLE, which is listed in Table~\ref{tab: runtime}. It can be found that the end-to-end runtime increase of UNiFi-DLE is in the scale of 1 ms as the number of antennas increases. On the contrary, the end-to-end runtime increase of the baseline PO-DLE+PA is in the scale of 100 ms. This originates from the SVD-based multi-cluster decoupling in UNiFi-DLE, which can be sped up with parallel computing, while the path extraction involving parameter estimation and clustering in PO-DLE+PA is much more time-consuming. Thus, the proposed UNiFi-DLE is more applicable for real-time channel extrapolation. 

\subsection{Sim-to-Real Experiments}
\label{subsec: sim2real}

To justify the unified generalizability of the proposed UNiFi-DLE in real-world scenarios, we consider the challenging sim-to-real experiments for further validation, which includes following channel measurement data in far-field and near-field scenarios. 
\begin{enumerate}
    \item \textbf{RENEW} \cite{tvt_du_2022_dataset}: This dataset is measured in ourdoor scenarios. The starting frequencies $f^{(\rm m)}_1$ and $f^{(\rm e)}_1$ are set as 2.402 GHz and 2.410 GHz, and the numbers of subcarriers are set as $K^{(\rm m)}=K^{(\rm e)}=26$. The subcarrier spacing is set as $\Delta f=0.3125$ MHz and $f^{(\rm e)}_1-f^{(\rm m)}_1=27\Delta f$ is held. A sub-array with 8 horizontal antennas and 4 vertical antennas is extracted, and the Rayleigh distance is 5 m. Since the horizontal distance between the user and the BS is larger than 15 m, the measured channel falls in a far-field scenario.  
    \item \textbf{ESPARGOS} \cite{dataset-espargos-0002}: This dataset is measured in an indoor near-field scenario. The starting frequencies $f^{(\rm m)}_1$ and $f^{(\rm e)}_1$ are set as 2.444 GHz and 2.453 GHz, and the numbers of subcarriers are set as $K^{(\rm m)}=K^{(\rm e)}=29$. The subcarrier spacing is set as $\Delta f=0.3125$ MHz and $f^{(\rm e)}_1-f^{(\rm m)}_1=29\Delta f$ is held. A UPA with 8 horizontal antennas and 4 vertical antennas serves as the BS, and the Rayleigh distance is 4.91 m. The distance between the user and BS ranges from 1.65 m to 6.76 m, and the measured channel falls in a near-field scenario. 
\end{enumerate}
In the sim-to-real experiments, we modify the number of subcarriers and the relative position of the measured and frequency bands in the simulated training datasets to align with the configuration of the real-world data. Then, the sim-to-real results are presented in Fig.~\ref{fig: sim2real}. It can be found that the proposed UNiFi-DLE can achieve the best sim-to-real performance in the RENEW LOS, ESPARGOS LOS, and ESPARGOS NLOS datasets, surpassing the PO-DLE+PA for $3\sim5$ dB. In the RENEW NLOS dataset, the proposed UNiFi-DLE can still achieve competitive generalizability compared to PO-DLE+PA. Specifically, in both LOS and NLOS scenarios of the near-field ESPARGOS datasets, the proposed UNiFi-DLE serves as the only solution to achieve the $\text{NMSE}<-6$ dB, while all other baselines fail. The rationale lies in two aspects. On the one hand, the spherical wave propagation significantly complicates the channel distribution in real-world and enlarges the sim-to-real gap, which severely degrades the alignment performance of PO-DLE+PA. On the other hand, the heterogeneous spherical wave propagation can be addressed by the physics-based disentanglement and alignment in UNiFi-DLE. Thus, the results in the real-world experiments precisely align with our theoretical analysis for unified near-field generalizability, which proves the practicality of proposed UNiFi-DLE to achieve unified generalization. 

\section{Conclusion}
Unified generalizability across near-field and far-field is critical for frequency-domain DL extrapolators in large-scale MIMO systems, which can significantly reduce the model deployment costs in diverse scenarios. In this paper, the unified generalization of DL extrapolators is achieved with strong physics interpretation. Firstly, we propose two key insights for unified generalization. On the one hand, we derive that heterogeneous angular profiles of near-field and far-field clusters cannot be effectively aligned. On the other hand, the delay profile is distance independent, which is unified across near-field and far-field scenarios. Secondly, the physics-based disentanglement and alignment framework is designed to address the generalizability challenges caused by diverse environments and heterogeneous propagation. This framework comprises three key components: multi-cluster decoupling, angle-delay disentanglement, and delay-domain alignment. Explicitly, the channel extrapolation is achieved by extrapolating the distributional stable aligned delay-domain features while directly reuse the heterogeneous angular-domain features. The physics-based interpretability can also be justified by the theoretical channel model and the real-world measurement data. Thirdly, the UNiFi-DLE is proposed for unified generalization and the workflow of training data preparation, model training, and inference. In the simulation and sim-to-real experiments, the proposed UNiFi-DLE is justified by comprehensive results comparing to the state-of-the-art, which achieves the best generalizability across diverse near-field and far-field scenarios. 
\appendices

\label{appdix: one-hop gbsm near-field}
\bibliographystyle{IEEEtran}
\bibliography{Ref}{}
\end{document}